\documentclass[12pt,letterpaper]{article}
\usepackage[utf8]{inputenc}

\usepackage[margin=1.25in]{geometry}
\usepackage{setspace}
\usepackage{titlesec}
\usepackage{tocloft}
\usepackage[hang,flushmargin]{footmisc}
\usepackage{appendix}

\usepackage{natbib}
\setcitestyle{authoryear, open={(}, close={)}}

\usepackage{graphicx}
\usepackage{grffile}
\usepackage{subcaption}
\usepackage{epsfig}
\usepackage{multicol}
\usepackage{listings}

\usepackage{booktabs}
\usepackage{dcolumn}
\usepackage{multirow}
\usepackage{hhline}
\usepackage{stackengine}
\usepackage{tablefootnote}
\usepackage{longtable}  
\usepackage{nicematrix} 

\usepackage[boxed,lined,linesnumbered]{algorithm2e}  

\usepackage{amsmath}
\usepackage{amssymb}
\usepackage{bbm}  
\usepackage{bm}   
\usepackage{optidef}  

\usepackage{microtype}
\usepackage{textcomp}
\usepackage[T1]{fontenc}  
\usepackage{lmodern}  

\usepackage{xspace}
\usepackage{color}
\usepackage[amsmath, thmmarks]{ntheorem}

\usepackage{pgf}
\usepackage{tikz} 
\usetikzlibrary{calc} 
\usetikzlibrary{arrows.meta} 
\usetikzlibrary{trees}

\usepackage[hidelinks]{hyperref} 

\DeclareMathOperator{\diag}{diag}

\DeclareMathOperator{\ci}{{CI}}
\DeclareMathOperator{\wci}{{WCI}}

\DeclareMathOperator{\bigO}{{\mathcal{O}}}

\DeclareMathAlphabet{\mathbfit}{OML}{cmm}{b}{it}

\DeclareMathOperator*{\argmin}{{argmin}}

\theoremseparator{.}
\newtheorem{theorem}{Theorem}
\theoremseparator{.}

\theoremseparator{.}

\newtheorem{lemma}[theorem]{Lemma}

\theorembodyfont{\normalfont}
\theoremseparator{.}
\theoremsymbol{\ensuremath{\diamond}}
\newtheorem{definition}{Definition}

\theoremstyle{nonumberplain}
\theorembodyfont{\normalfont}
\theoremseparator{.}
\theoremsymbol{\ensuremath{\diamond}}

\theoremstyle{nonumberplain}
\theoremheaderfont{\itshape}
\theorembodyfont{\normalfont}
\theoremseparator{.}
\theoremsymbol{\ensuremath{\blacksquare}}

\newcommand{\R}{\mathbb{R}}

\newcommand{\s}[1]{{\left\{#1\right\}}}
\newcommand{\p}[1]{{\left(#1\right)}}

\bmdefine{\bone}{1}
\bmdefine{\bzero}{0}

\newcommand{\br}{\bm{r}}

\newcommand{\bx}{\bm{x}}

\newcommand{\Db}{\mathbf{D}}

\newcommand{\Ib}{\mathbf{I}}

\newcommand{\Pb}{\mathbf{P}}

\newcommand{\Xb}{\mathbf{X}}

\newcommand{\bmu}{\bm{\mu}}

\title{\bf Powerful significance testing for unbalanced clusters}

\author{%
	Thomas H.\ Keefe\thanks{
		Department of Statistics \& Operations Research,
		UNC-Chapel Hill. \texttt{tkeefe}\texttt{@live.unc.edu}}
	\and J.\ S.\ Marron\thanks{
		Department of Statistics \& Operations Research,
		UNC-Chapel Hill.
		\texttt{marron}\texttt{@unc.edu}}
}

\date{August 24, 2023}

\begin{document}

\maketitle

\begin{abstract}
\noindent Clustering methods are popular for revealing structure in data, particularly in the high-dimensional setting common to contemporary data science.
A central \textit{statistical} question is, ``are the clusters really there?''
One pioneering method in statistical cluster validation is \textit{SigClust}, but it is severely underpowered in the important setting where the candidate clusters have unbalanced sizes, such as in rare subtypes of disease.
We show why this is the case, and propose a remedy that is powerful in both the unbalanced and balanced settings, using a novel generalization of $k$-means clustering.
We illustrate the value of our method using a high-dimensional dataset of gene expression in kidney cancer patients.
A Python implementation is available at
\url{https://github.com/thomaskeefe/sigclust}.
\end{abstract}

\section{Introduction}
\label{sec:sigclust:introduction}
Clustering is a rich topic that finds broad application, such as in bioinformatics, communication, and business.
The ability to collect vast quantities of biological data has led to much success in using clustering methods in bioinformatics to investigate disease subtypes, see e.g., the celebrated paper by \cite{perouMolecularPortraitsHuman2000}.
Finding clusters is typically an exploratory process that is followed up with validation by experts in the underlying biology.
Despite a large number of available clustering algorithms, many fewer are available for statistically validating the results of clustering.
Validation is typically performed using either \textit{internal measures}, which concern cluster cohesion and separation, or \textit{external measures}, which compare the clustering to some known classification.
\cite{halkidi_vazirgiannis_hennig_2015} and \cite{meila_external} provide overviews of internal and external measures respectively.
However, neither are sufficient for disease subtyping: the internal measures lack statistical guarantees, and the external measures cannot be applied because typically there is not a known classification to which to compare the results.
\textit{Statistical} procedures to validate clustering include SigClust \citep{sigclust1},
which tests whether two clusters are ``really there'' by determining if they produce a stronger cluster index (CI, an internal validation measure) than could be found under a hypothesis of just one cluster.
SigClust has proved popular in bioinformatics and has extensions for high-dimensional data \citep{sigclust2} and hierarchical clustering \citep{sigfuge}.
Other popular approaches include the \textit{gap statistic} \citep{gapstat}, which is an estimator for the true number of clusters,
and \textit{consensus clustering} \citep{monti}, which is based on resampling.
\textit{Bayesian mixture modeling} \citep[Chapter~20]{BDA} offers a Bayesian approach.
Most recently, the \textit{RIFT} test \citep{rift} has outperformed SigClust in data with a large second eigenvalue.

This paper addresses a major limitation of SigClust, which is that its statistical power is severely limited when the clusters have very unbalanced sizes.
In clinical datasets, such as the kidney cancer example in this work, this may cause SigClust to fail to validate important clusters representing rare subtypes of a disease.
The underlying reasons have to do with the fact that SigClust is built on ideas from $k$-means clustering, and the observation, perhaps first made by \cite{macqueen1967}, that $k$-means prefers to produce \textit{balanced} clusterings;
see also \cite{mirkin_kmeans}.
We elucidate this issue using the geometry of $k$-means, and discuss why it leads SigClust to fail in this setting. We then propose an improvement, \textit{Weighted SigClust}, using a novel generalization of $k$-means clustering that better recovers unbalanced clusters.

The paper is organized as follows.
In Section \ref{sec:sigclust:sigclustreview} we review SigClust and motivate our work using a kidney cancer dataset with a potential rare subtype that SigClust is not powered to detect.
In Section \ref{sec:sigclust:proposal} we present Weighted SigClust and use it to find support for the rare subtype.
Section \ref{sec:sigclust:optimization} discusses an algorithmic implementation.
Section \ref{sec:sigclust:discussion}
is discussion.

\subsection{Notation} \label{sec:sigclust:notation}
\newcommand{\bxbar}{\bm{\bar{x}}}
Throughout this work we will assume a dataset taking the form $\s{\bx_1, \cdots, \bx_n} \subset \R^d$, alternately denoted by the matrix $\Xb \in \R^{d\times n}$.
When $\Xb$ is partitioned into two clusters, we will denote them by the pair $(C_1, C_2) \in \mathcal{P}_n$, where $\mathcal{P}_n$ denotes the collection of two-set partitions of the indices $\s{1, \cdots, n}$:
\begin{equation}\label{eq:Pn}
	\mathcal{P}_n = \s{(C_1, C_2) \,:\, C_1 \uplus C_2 = \s{1, \cdots, n}}.
\end{equation}
The cluster means, or \textit{centroids}, are denoted as $\bxbar^\p{1}$ and $\bxbar^\p{2}$, and the overall data mean veftor as $\bxbar$.

\section{Review of the cluster index and SigClust} \label{sec:sigclust:sigclustreview}
In this section we review the SigClust procedure, which tests the validity of a given $k=2$ clustering.
It relies on a heuristic measure of cluster strength called the cluster index,
which we review in Section \ref{sec:sigclust:sigclustreview:ci}.
We continue with a review of SigClust in Section \ref{sec:sigclust:sigclustreview:sigclust}.

\subsection{The cluster index} \label{sec:sigclust:sigclustreview:ci}
The \textit{cluster index} (CI), so named by \cite{sigclust1}, is a heuristic measure of the strength of a $k=2$ clustering.
For a dataset $\Xb$ and clusters $(C_1, C_2)$,
(recall the notation established in Section \ref{sec:sigclust:notation}),
the CI is defined
\begin{equation}\label{eq:CI}
\ci(C_1, C_2) = \frac{\sum_{i \in C_1} ||\bx_i - \bxbar^\p{1} ||^2 + \sum_{i \in C_2} ||\bx_i - \bxbar^\p{2}||^2 }{\sum_{i=1}^n ||\bx_i - \bxbar||^2} .
\end{equation}
The numerator of (\ref{eq:CI}) is called the \textit{within-cluster sum of squares}, and is the typical objective function of $k$-means clustering. See \cite{steinley_k-means_2006} for a comprehensive review of $k$-means.
The denominator of (\ref{eq:CI}) is called the \textit{total sum of squares}, and has the effect of standardizing the CI to the unit interval.

The value of the CI comes from the following observation.
If we call two groups of points \textit{clusters}, then there are at least two ways to make the clusters ``stronger'':
\begin{enumerate}
	\item Tighten the spread of the points around the cluster centers, e.g. by decreasing the within-cluster sums of squares, or,
	\item Move the clusters farther apart, e.g., by increasing the distances between the centroids.
\end{enumerate}
The CI is a useful measure of cluster strength because it decreases by doing either of the two above.

From the perspective of $k$-means clustering, minimizing (\ref{eq:CI}) versus just minimizing its numerator are equivalent, because the denominator is constant in the data.
The purpose of including the denominator is to give the criterion the upper bound of 1.
The CI achieves this upper bound iff the clusters are moved so close together that their centroids collide, reflecting the notion that there can be no ``weaker'' clusters than these.

On the other hand, the CI reaches its lower bound of 0 iff the clusters are so tight that the points are exactly piled on the cluster centroids.
As long as the centroids are themselves separated, the CI of these clusters will be 0, and this neatly reflects the notion that there can be no ``stronger'' clusters than these.
We design the methods in this paper to retain these properties, which we encapsulate in the following definition.

\begin{definition}[The CI property]\label{def:ciproperty}
	A clustering criterion $f: (\R^{d\times n} \times \mathcal{P}_n) \to [0,1]$
	has \textit{the CI property} iff
	\begin{equation} \label{eq:ciproperty1}
	f(\Xb, C_1, C_2) = 0 \iff
	\sum_{i \in C_1} ||\bx_i - \bxbar^\p{1}||^2
	=
	\sum_{i \in C_2} ||\bx_i - \bxbar^\p{2}||^2
	= 0,
	\end{equation}
	and
	\begin{equation} \label{eq:ciproperty2}
		f(\Xb, C_1, C_2) = 1 \iff
		\bxbar^\p{1} = \bxbar^\p{2}
		\text{ and }
		\sum_{i=1}^n ||\bx_i - \bxbar||^2
		> 0.
	\end{equation}
\end{definition}
The discussion above may be summarized by saying that the numerator of (\ref{eq:CI}) provides (\ref{eq:ciproperty1}), while the denominator provides (\ref{eq:ciproperty2}).

\subsection{The SigClust procedure} \label{sec:sigclust:sigclustreview:sigclust}

\begin{figure}[h!]
	\centering
	\includegraphics[width=6in]{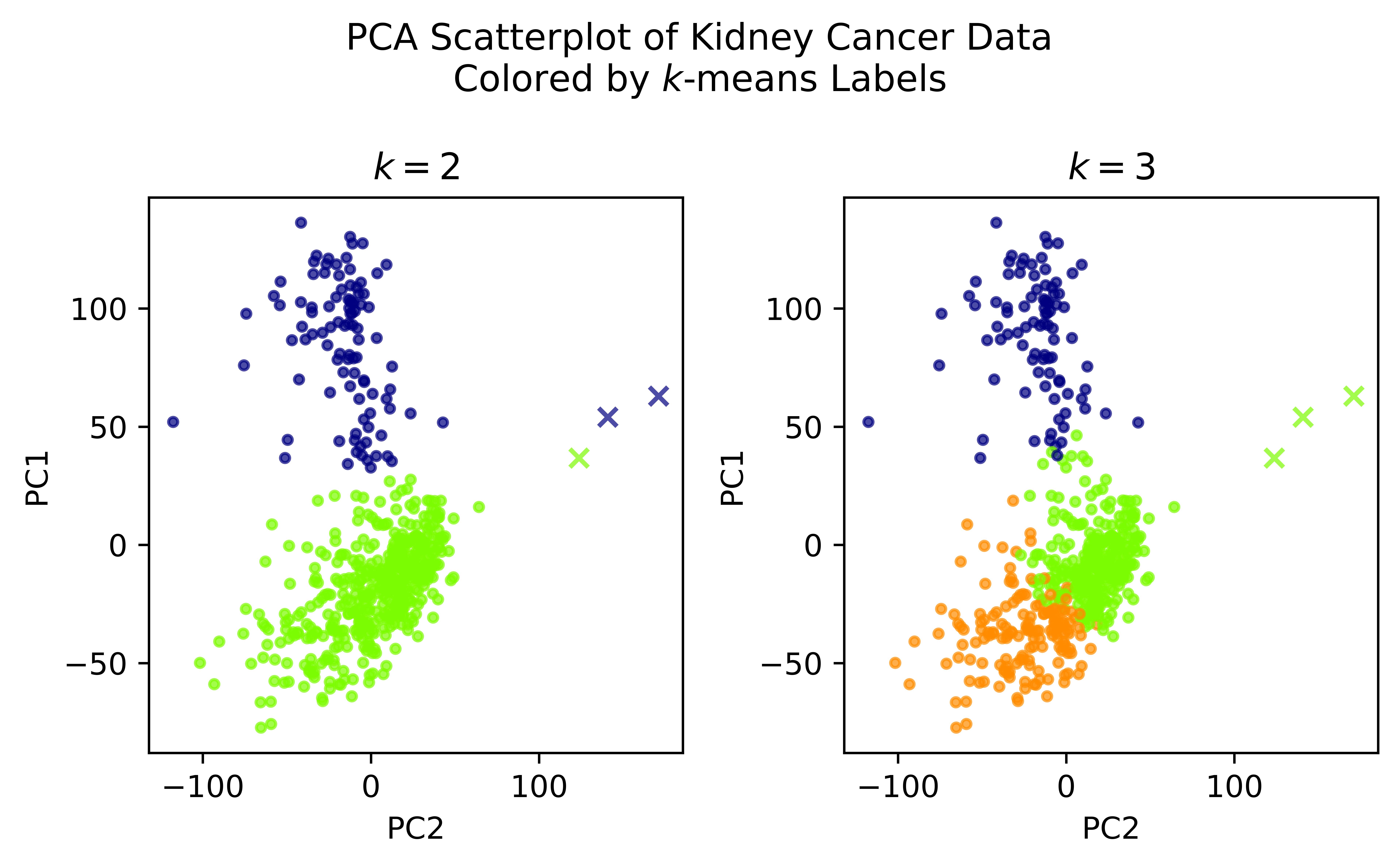}
	\caption{PC1-PC2 scatterplot of kidney cancer data, colored by $k$-means labels. There is a clear outlier cluster marked in crosses, but $k$-means clustering will not select it with either $k=2$ (blue/green at left) or $k=3$ (blue/green/orange at right).}
	\label{fig:kidney_pca}
\end{figure}

We review SigClust using an example from the PanCan database of The Cancer Genome Atlas \citep[TCGA;][]{tcga}.
The particular dataset we use here is PanCan's \textit{Kidney Cancer Data},
which contains gene expression measurements of $n=551$ kidney tumors measured across $d=12,478$ genes.
Figure \ref{fig:kidney_pca} shows the first two principal components of this dataset, colored by the $k$-means labels with both $k=2$ (left) and $k=3$ (right).
Visually, the clustering produced by $k=2$ is quite sensible, but there is a clear outlier group of three points marked with crosses.
Could they represent a rare subtype of kidney cancer?
One might think that simply increasing $k$ to 3 would separate this outlier group as the third cluster,
but as the right panel shows, $k=3$ results in splitting the large cluster in the bottom half of the figure instead.
We therefore consider two interesting clusterings of this dataset:
\begin{enumerate}
	\item \texttt{(2-means)} the 2-means clustering, blue versus green in the left panel of Figure \ref{fig:kidney_pca}
	\item \texttt{(outlier-inlier)} the outlier group versus the rest, crosses versus dots in Figure \ref{fig:kidney_pca},
\end{enumerate}
which we will test using SigClust.

Following \citet[Section~13.2]{oodabook},
SigClust can be used in two modes:
\begin{definition}[Exploratory/confirmatory modes]\label{def:sigclustmodes}
	Applying SigClust to a data sample $\Xb$ with a particular clustering of interest, $(C_1, C_2) \in \mathcal{P}_n$, is called \textit{confirmatory mode} SigClust.
	Applying SigClust to $\Xb$ \textit{without} a particular clustering in mind is referred to as \textit{exploratory mode} SigClust.
	When used in exploratory mode, SigClust uses the labels from 2-means clustering as the clustering of interest.
	Exploratory mode is therefore equivalent to using confirmatory mode with 2-means labels.
\end{definition}
For our kidney cancer example, \texttt{2-means} refers to using SigClust
in exploratory mode, and \texttt{outlier-inlier} to confirmatory mode.

SigClust assesses whether the CI of a candidate clustering is smaller, i.e., stronger, than could be expected if the population contained only one cluster.
To do this, SigClust must define appropriate null and alternative hypotheses, and then compute or estimate the distribution of the minimal CI under the null.

SigClust defines a \textit{cluster} as a group of observations from a single multivariate Gaussian distribution (which may be spherical or stretched).
This motivates the hypotheses
\begin{itemize}
	\item $H_0$: the distribution is a multivariate Gaussian, i.e., has only one cluster;
	\item $H_A$: the distribution is not a single multivariate Gaussian.
\end{itemize}

It is important to note that $H_A$ takes the form ``not $H_0$'' rather than ``a mixture of two multivariate Gaussians,'' or even ``the population has clusters.'' Thus, $H_0$ and $H_A$ cover a quite general data model.
In the strictest sense, SigClust is a test of Gaussianity, but one that is powerful against $H_0$ when the distribution does in fact support clusters.
Its power comes from using the CI as its test statistic, which is small when the data have two clusters,
and large when they constitute one cluster.

SigClust estimates the null distribution of the CI using a parametric bootstrap scheme.
A large number of synthetic datasets, say 100 or 1,000, each with $n$ observations, are simulated using a $d$-dimensional Gaussian.
Each synthetic dataset is clustered by 2-means clustering, and the resulting CI is then a parametric bootstrap sample from the null distribution.

In low-dimensional settings, these synthetic datasets can be simulated from the maximum-likelihood Gaussian, $N_d( \bxbar, \bm{\hat{\Sigma}})$.
However, since the CI is invariant to translation and rotation, it is more statistically efficient to sample from
$N_d(\bzero, \diag (\hat{\lambda}_1, \cdots, \hat{\lambda}_d))$, where $\hat{\lambda}_i$ are estimates of the eigenvalues of the covariance matrix.
\cite{sigclust2} showed that in high-dimensional settings, a soft-thresholding approach improves the estimation of these eigenvalues, which improves SigClust's power.

Finally, if the candidate CI is unusually small with respect to the simulated null distribution, SigClust concludes that the candidate clustering is stronger than would be plausible under $H_0$.
The significance may be assessed by computing an empirical p-value for the candidate CI with respect to the left tail of the simulated null distribution.

\begin{figure}[h!]
	\centering
	\includegraphics[width=.6\linewidth]{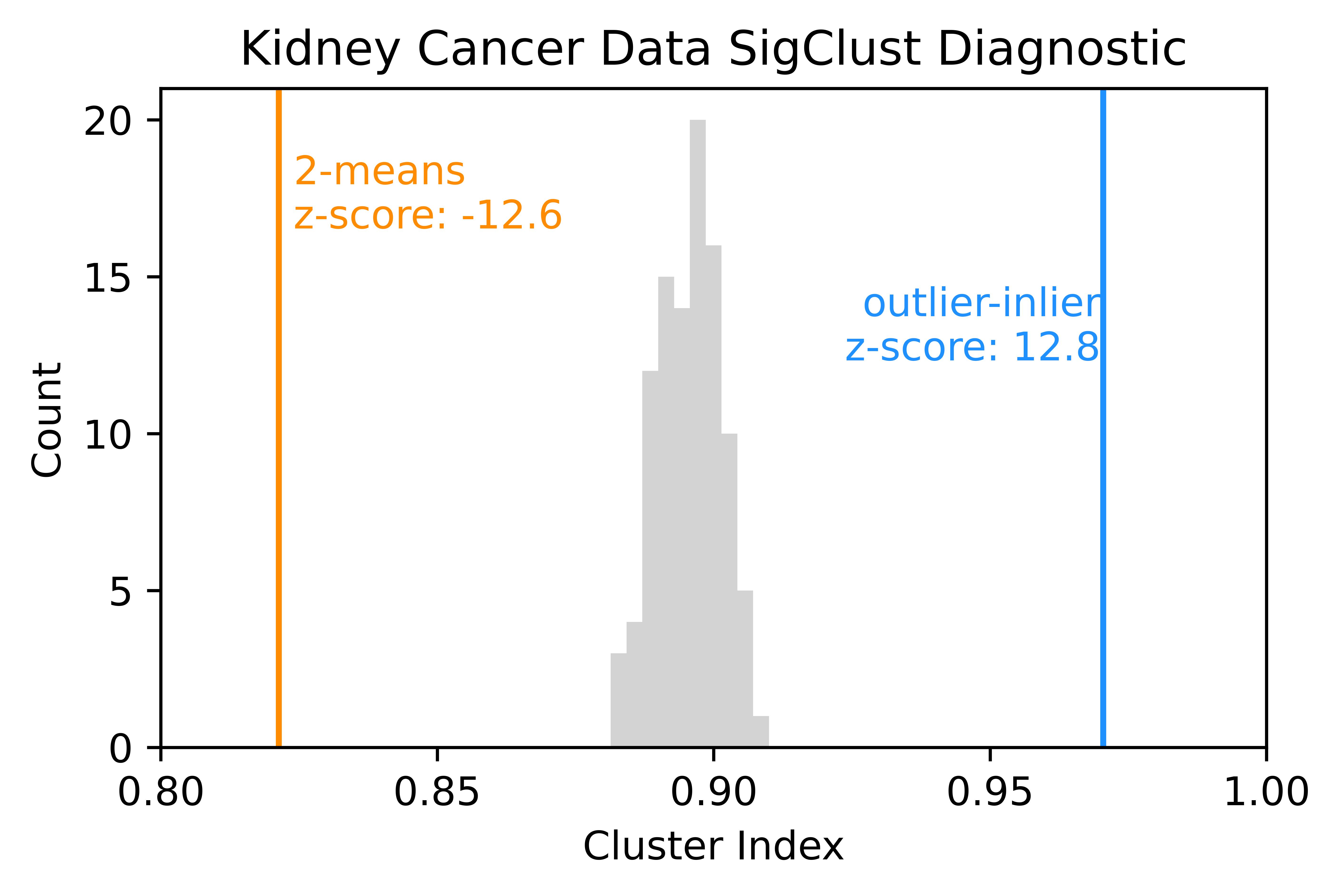}
	\caption{SigClust diagnostic comparing the \texttt{2-means} and \texttt{outlier-inlier} clusterings. The histogram shows the simulated distribution of the CI under $H_0$; the vertical lines show the CIs of our two clusterings. Figure shows that conventional SigClust is unable to validate the \texttt{outlier-inlier} clustering.}
	\label{fig:kidney_sigclust_conventional}
\end{figure}

We now apply SigClust to the two clusterings of interest from the example in the left panel of Figure \ref{fig:kidney_pca}: \texttt{2-means}, which is shown with blue and green colors; and \texttt{outlier-inlier}, which is shown using crosses and dots.
The results are presented using the \textit{SigClust diagnostic plot} (Figure \ref{fig:kidney_sigclust_conventional}).
The simulated null distribution is shown with a histogram, and the candidate CIs of interest are shown with vertical lines for comparison.
First, we discuss the \texttt{2-means} clustering.
We see that the \texttt{2-means} CI is lower (stronger) than any of the 100 CIs simulated under $H_0$, which would produce an empirical p-value of less than $\frac{1}{100 + 1}$.
While this empirical p-value allows us to reject $H_0$ at level .01, we typically also report a \textit{z-score}, which is computed by standardizing the candidate CI with respect to the mean and standard deviation of the simulated null distribution.
In this case, the \texttt{2-means} CI is 11.8 standard deviations lower than the mean CI under $H_0$.
The z-score is helpful when comparing multiple candidate CIs, which may have all the same empirical p-value.
However, using either the z-score or the empirical p-value, we can conclude that the \texttt{2-means} clusters are much stronger than should occur in unclustered, Gaussian data.

The other CI in Figure \ref{fig:kidney_sigclust_conventional} is the one produced by the \texttt{outlier-inlier} clustering, which exemplifies the major point of this paper.
There are two things we want to say about it.
First, it is near 1, saying that this clustering is weak with respect to the CI criterion.
We will argue that this criterion is not fair to unbalanced clusters.
Second, it is far outside the range of CIs produced under $H_0$, but in the right tail of the null distribution instead of the left.
SigClust is defined as a left-tail test because the CI is small when clusters are present; placing the rejection region in the left tail therefore maximizes power.
Furthermore, although this CI is far outside the range of those produced under $H_0$, there is no useful conclusion available.
In fact, any dataset can be labeled to produce a CI out in the right tail, regardless of whether there is cluster structure or not.
For example, simply assigning points to clusters by independent coin flips tends to produce overlapping clusters whose means are very close to each other, and to the overall mean.
Hence, the resulting CI will be close to 1 and larger than any of the CIs in the SigClust null distribution, but clearly no clustering structure has been assessed.

Although the \texttt{outlier-inlier} clustering is weak with respect to the CI criterion, it
looks like a clustering of keen interest in the PCA scatterplot (Figure \ref{fig:kidney_pca}),
suggesting the CI is not the appropriate measure for this scenario.
In fact, the large CI occurs specifically because the clusters are so unbalanced.
The three outlier points account for almost none of the of the total variance of the 551 observations, and barely influence the mean.
Therefore, the CI is essentially comparing the total variance of the 548 inlier points around \textit{their} mean to the total variance of all 551 points around the very nearby \textit{overall} mean.
Hence the numerator and denominator of the CI are nearly the same, giving a value near 1.
In general, unbalanced clusters will have large CIs for this reason.
Another example is visually explained in Figure \ref{fig:hotdogtruevs2meanssquares} in Section \ref{sec:sigclust:wci}.
Unbalanced clusters therefore require a criterion that accounts for the relative sizes.

The \texttt{outlier-inlier} clustering provides just one example of unbalanced clusters that may be critically important, but that SigClust is not powered to validate.
In the following section, we generalize the CI to give more weight to small clusters, so that the \texttt{outlier-inlier} clustering is strong.
Integrating this modified clustering objective into SigClust greatly improves its power when clusters are unbalanced,
and allows SigClust to recognize the \texttt{outlier-inlier} clustering as strongly significant.

\section{The proposed methodology} \label{sec:sigclust:proposal}
SigClust lacks power in the case of strongly unbalanced clusters because such clusters produce large CIs, as discussed in Section \ref{sec:sigclust:sigclustreview:sigclust}.
To address this issue, we propose a modification of the
CI, called the \textit{weighted CI} (WCI), that can recognize unbalanced clusters as strong.
Using the WCI as SigClust's test statistic then rectifies SigClust's power in the unbalanced setting.
In Section \ref{sec:sigclust:wci} we motivate and develop the WCI using a toy dataset, and propose a clustering method around it, called \textit{WCI clustering}.
In Section \ref{sec:sigclust:wcisigclust} we integrate WCI clustering into SigClust, which we call \textit{Weighted SigClust}, and present toy examples showing its improved power over conventional SigClust.
Finally, we return to the kidney cancer example, where we show that Weighted SigClust finds strong support for both clusterings of interest.

\subsection{The weighted cluster index}
\label{sec:sigclust:wci}
In this section we propose the \textit{weighted cluster index} (WCI), which weights the sums-of-squares in (\ref{eq:CI}) by a power of the cluster sizes.
The WCI function is:
\begin{equation}\label{eq:wci}
	\wci_g(C_1, C_2) =
	\frac{
		\frac{1}{|C_1|^g} \sum_{i \in C_1} || \bx_i - \bxbar^\p{1} ||^2
		+
		\frac{1}{|C_2|^g} \sum_{i \in C_2} || \bx_i - \bxbar^\p{2} ||^2
	}
	{
		\frac{1}{|C_1|^g} \sum_{i \in C_1} || \bx_i - \bxbar ||^2
			+
		\frac{1}{|C_2|^g} \sum_{i \in C_2} || \bx_i - \bxbar||^2
	},
\end{equation}
where the exponent $g$ is a tuning parameter that controls how much weighting is applied.

The idea of the WCI is to weight the sums of squares so as to allow small clusters to matter more in the optimization, while retaining the CI property (Definition \ref{def:ciproperty}).
It is easy to verify that that WCI has the CI property; furthermore,
when the clusters $C_1$ and $C_2$ have the same size, the WCI is equivalent to the CI.
When $g=0$ we recover the original CI.
When using WCI clustering in SigClust, we recommend trying each of $g = 0,$ $0.25$, and $0.5$, an using the value with the strongest z-score.

\begin{figure}[htbp]
	\centering
	\includegraphics[width=12cm]{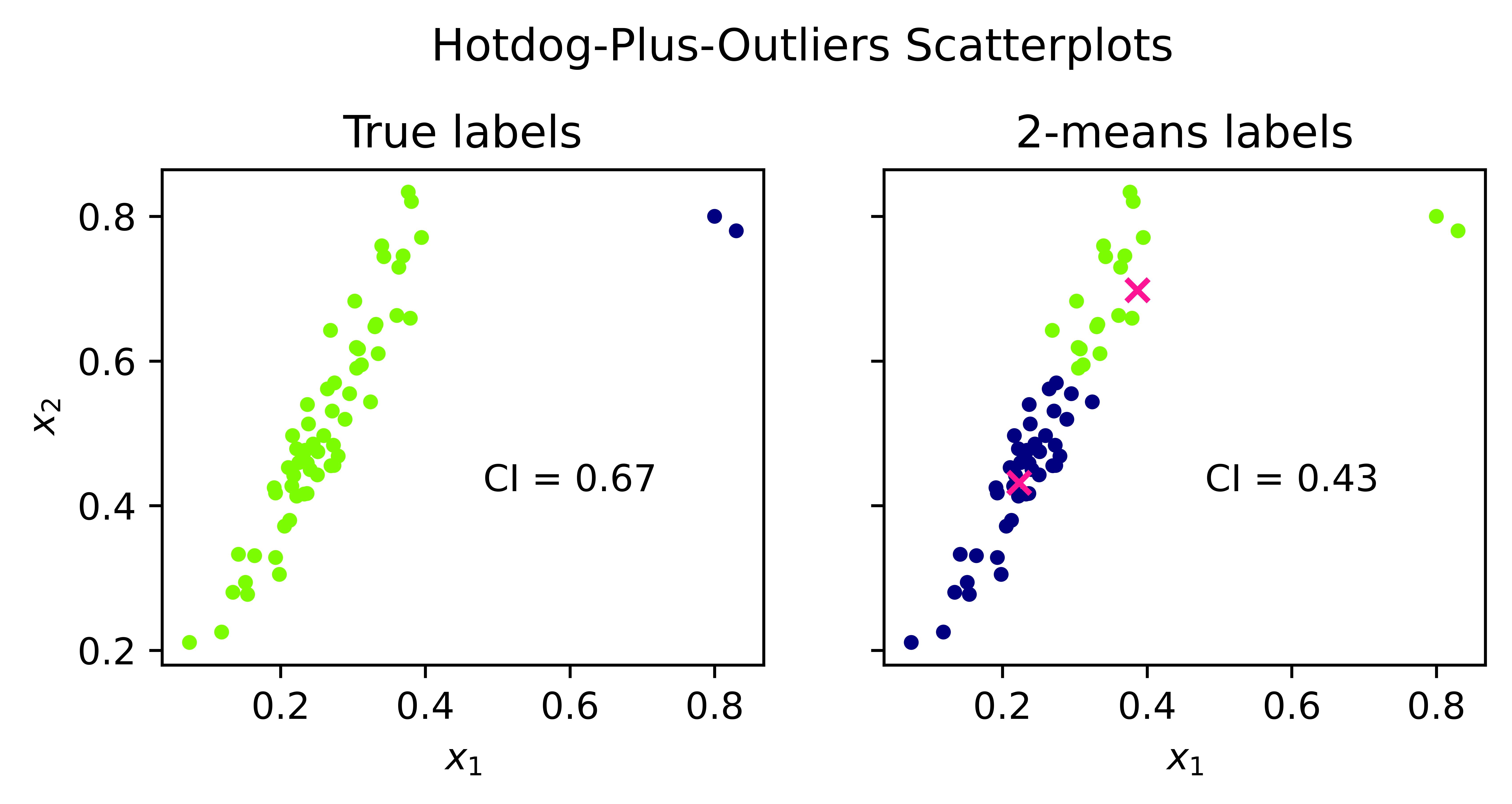}
	\caption{Hotdog-Plus-Outliers toy dataset, colored by true class labels (left) and $2$-means labels (right). $2$-means clustering does not recover the true labels, because breaking apart the hotdog produces a much lower CI of 0.43. The pink crosses in the right panel indicate the $2$-means centroids.}
	\label{fig:hotdog}
\end{figure}

\begin{figure}[htbp]
	\centering
	\includegraphics[width=0.7\linewidth]{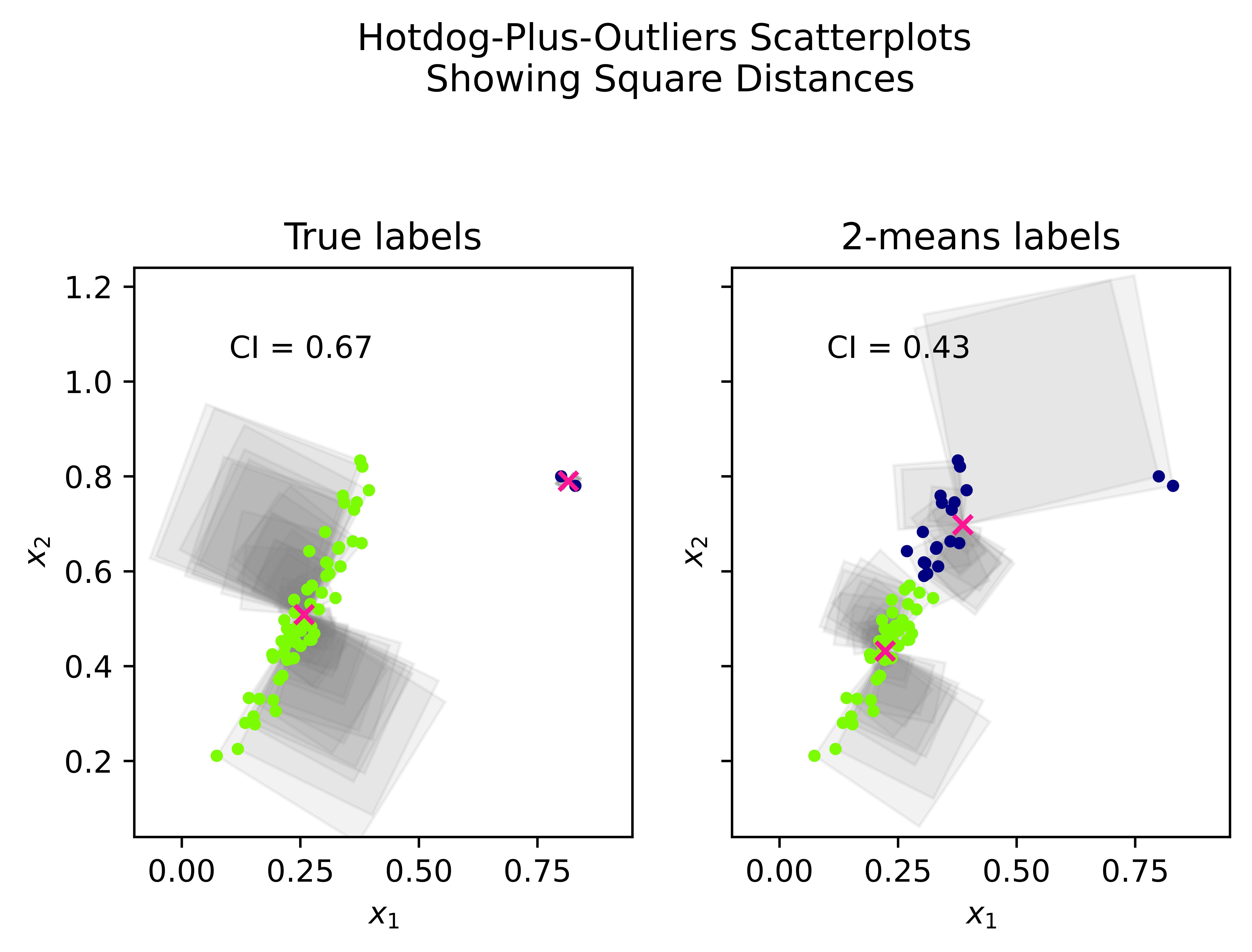}
	\caption{Hotdog-Plus-Outliers dataset colored by true labels (left) and 2-means labels (right). The semitransparent gray squares show the squared distances to the centroids (pink crosses). The amount of gray gives a visual indication of within-cluster sums of squares. The greater accumulation of gray on visualizes the difference between the sums-of-squares.}
	\label{fig:hotdogtruevs2meanssquares}
\end{figure}

We motivate the WCI using a toy 2-D dataset, \textit{Hotdog-Plus-Outliers}, which is a subset of the \textit{Four Clusters} dataset from \citet[Chapter~12]{oodabook}.
A scatterplot is shown in Figure \ref{fig:hotdog}.
These data are simulated from two unbalanced clusters: a 60-point stretched Gaussian ``hotdog,'' and two outlier points.
We call this labeling the \textit{true labels}.
The left panel of Figure \ref{fig:hotdog} colors the points by the true labels, while the right panel colors them by the 2-means labels.
The 2-means clustering prefers to break apart the large cluster rather than recover the true labels.
The true labels have a CI of 0.67, while the 2-means labels have a much stronger CI of 0.43.
The reason that $2$-means does not recover the correct clusters is that the two outlier points are just too few to contribute much to the optimization.
Figure \ref{fig:hotdogtruevs2meanssquares} provides a geometric explanation by visualizing each squared distance as a gray square, where one edge of the square connects a point to its associated centroid.
The true labels achieve two tiny squares on the outlier points by partitioning them into their own cluster, but at the expense of quite a few large squares in the hotdog cluster.
The 2-means labels, however, break apart the hotdog.
The semitransparent squares produce darker areas of gray when they overlap, so the total amount and darkness of gray gives a visual indication of the within-cluster sum of squares for each labeling.
The greater total accumulation of gray in the left panel compared to the right
illustrates that it is worth trading those two tiny squares on the blues in order to get more small and medium squares on the hotdog.
The reason that $k$-means often does not recover unbalanced clusters is that it is just not worth getting a few tiny squares on small clusters, when large clusters can be broken apart to get a larger number of small and medium-sized squares.
Therefore, our approach in defining the WCI
is to use the sizes of the clusters as weights,
which allows small clusters to play a larger role in the optimization.

\begin{figure}[htbp]
	\centering
	\includegraphics[width=12cm]{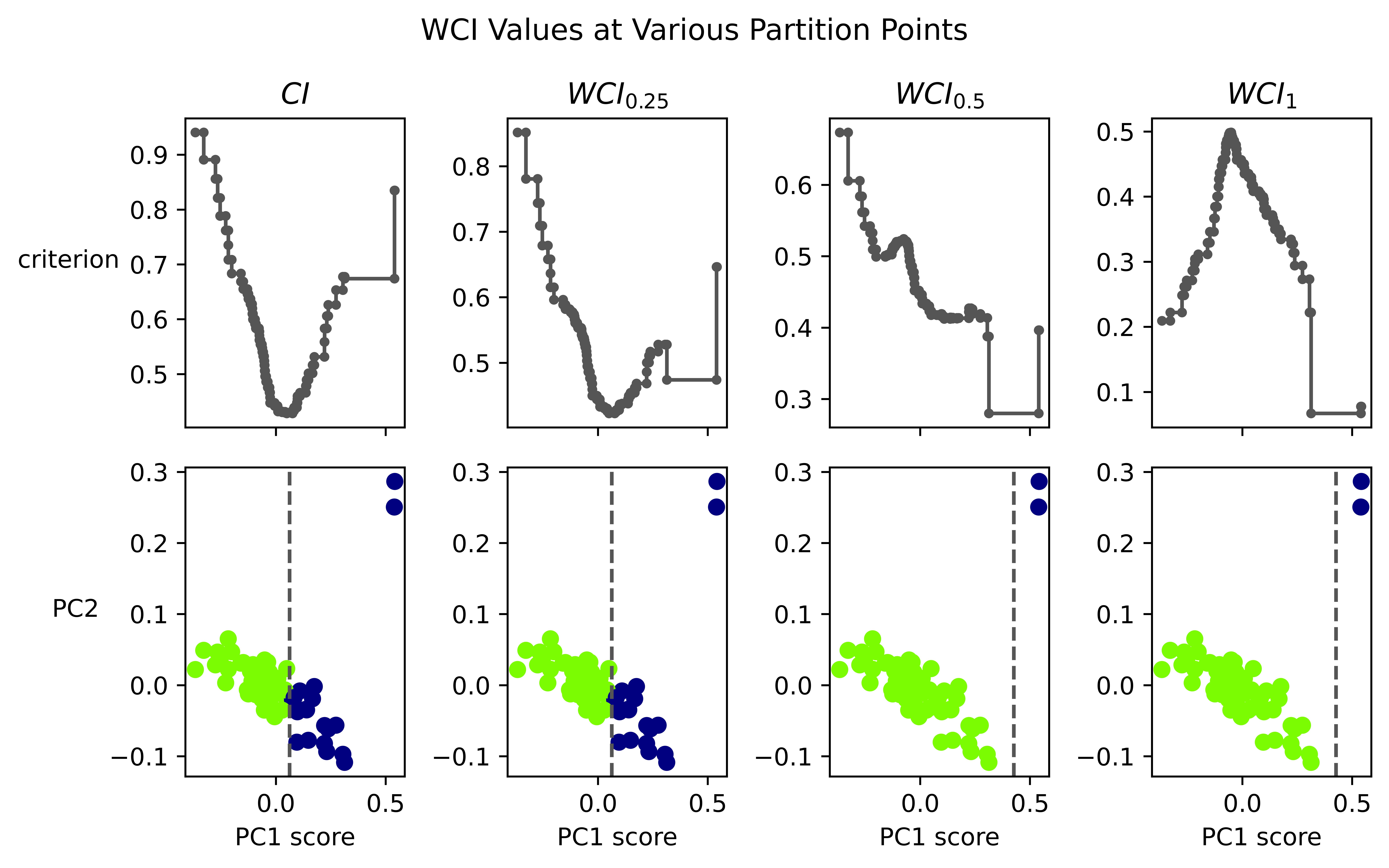}
	\caption{Bottom row: PC1-2 scatterplots of Hotdog-Plus-Outliers dataset. The dashed lines indicate the minimum of the criterion in question. Top row: CI, $\wci_{0.25}$, $\wci_{0.5}$, and $\wci_{1}$ evaluated at different partitions along PC1.}
	\label{fig:compare_cis_on_hotdog}
\end{figure}

In Figure \ref{fig:compare_cis_on_hotdog}
we rotate this dataset to its principal axes and consider the clusters generated by sliding a partition line along PC1.
We then plot the original CI, $\wci_{0.25}$, $\wci_{0.5}$, and $\wci_{1}$ as functions of the partition point on PC1.
The dashed vertical line marks the partition that minimizes the criterion.
We see that conventional CI and $\wci_{0.25}$ are minimized by
splitting the hotdog roughly in half, but $\wci_{0.25}$ is
also nearly minimized by separating the outliers from the hotdog.
It therefore considers both clusterings to be good.
$\wci_{0.5}$ and $\wci_{1}$ are both minimized by separating the outliers from the hotdog, but $\wci_{0.5}$ also indicates that splitting the hotdog in two is reasonable.
Although $\wci_{1}$ has a nice interpretation of minimizing the combined mean squares in each cluster, we do not recommend using it.
In practice, $\wci_{1}$ tends to pluck off one or two outliers, regardless of the cluster structure.

\begin{figure}[htbp]
	\centering
	\textbf{Scatterplot Matrix of Cluster Sizes for Various Choices of $g$ and $d$}\\
	\textbf{One point = one simulation}
	\vskip
	\baselineskip
	\begin{subfigure}{.45\textwidth}
		\centering
		\includegraphics[width=3in]{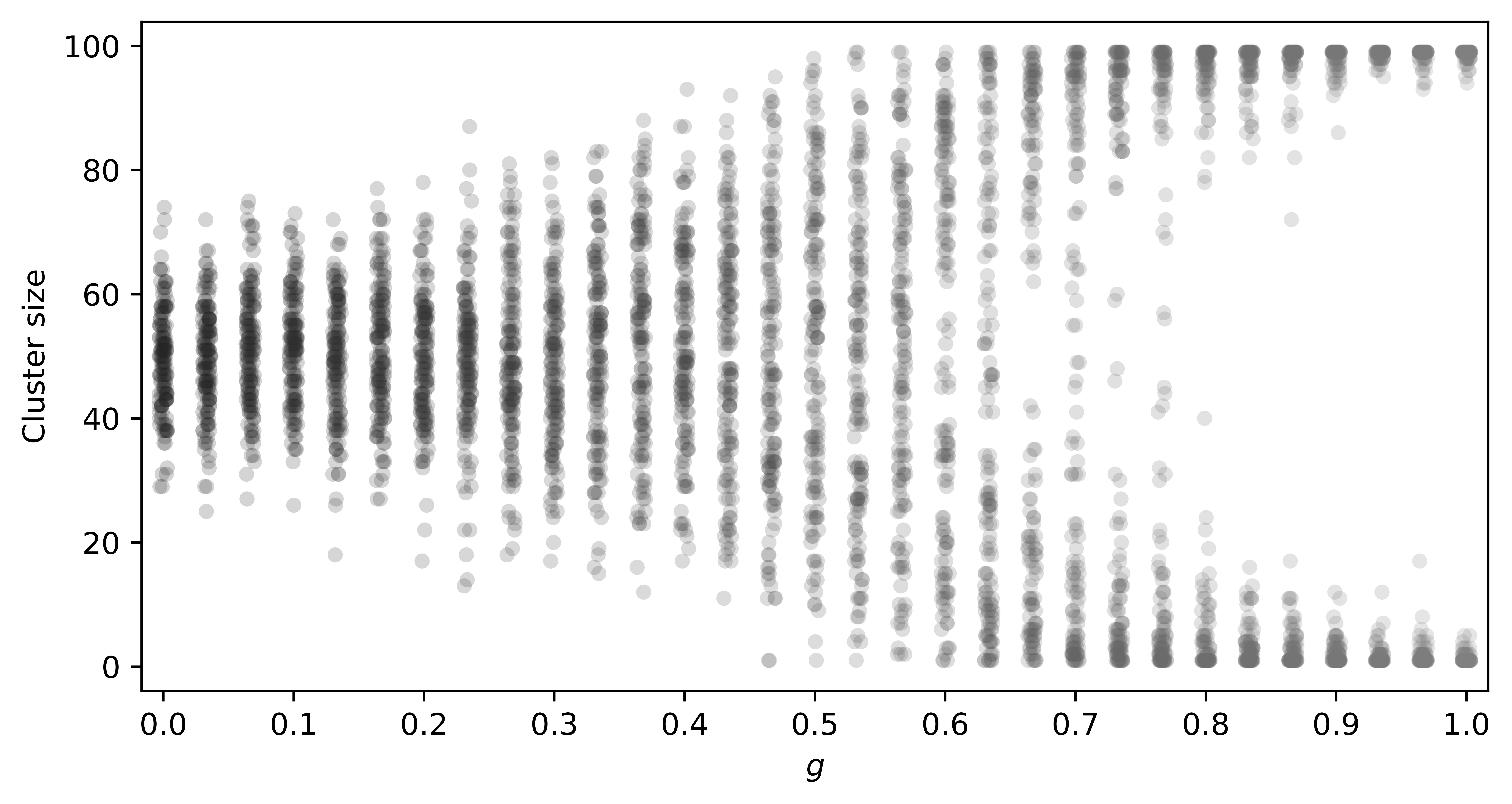}
		\caption{$d=1$}
		\label{fig:ScatterRandomClusterSizesD1}
	\end{subfigure}
	\hfill
	\begin{subfigure}{.45\textwidth}
		\centering
		\includegraphics[width=3in]{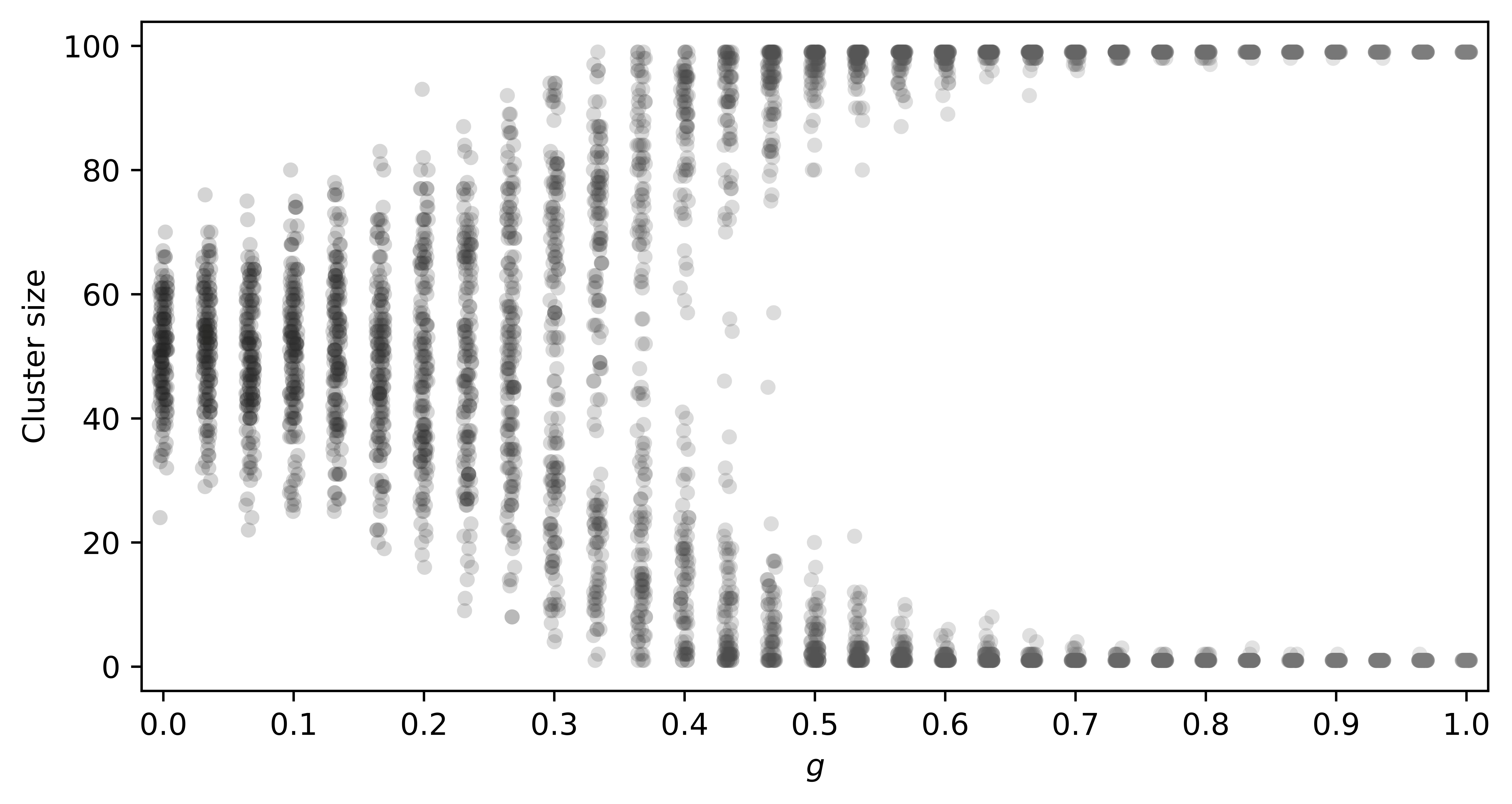}
		\caption{$d=4$}
		\label{fig:ScatterRandomClusterSizesD4}
	\end{subfigure}
	\vskip\baselineskip
	\begin{subfigure}{.45\textwidth}
		\centering
		\includegraphics[width=3in]{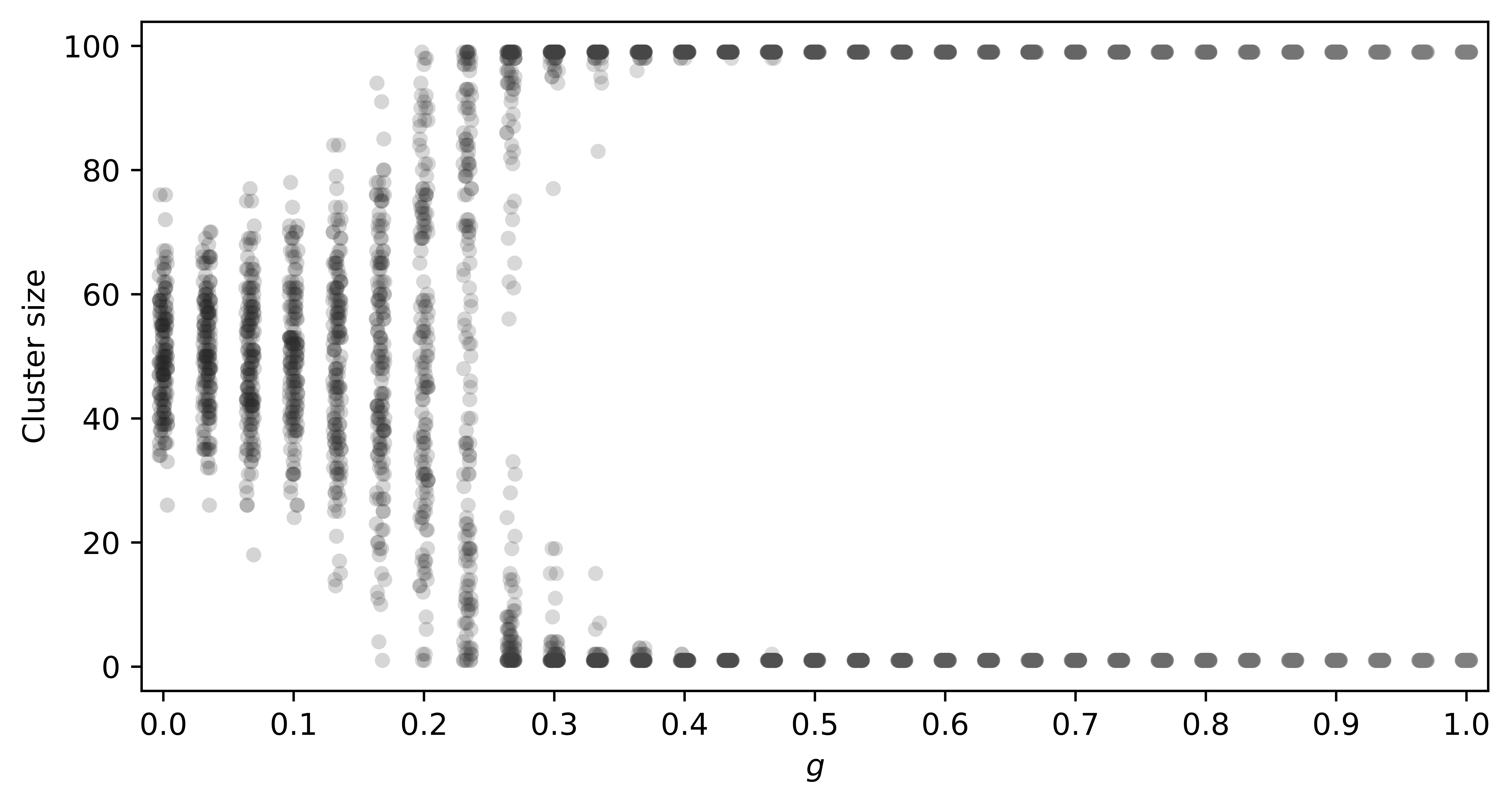}
		\caption{$d=32$}
		\label{fig:ScatterRandomClusterSizesD32}
	\end{subfigure}
	\hfill
	\begin{subfigure}{.45\textwidth}
		\centering
		\includegraphics[width=3in]{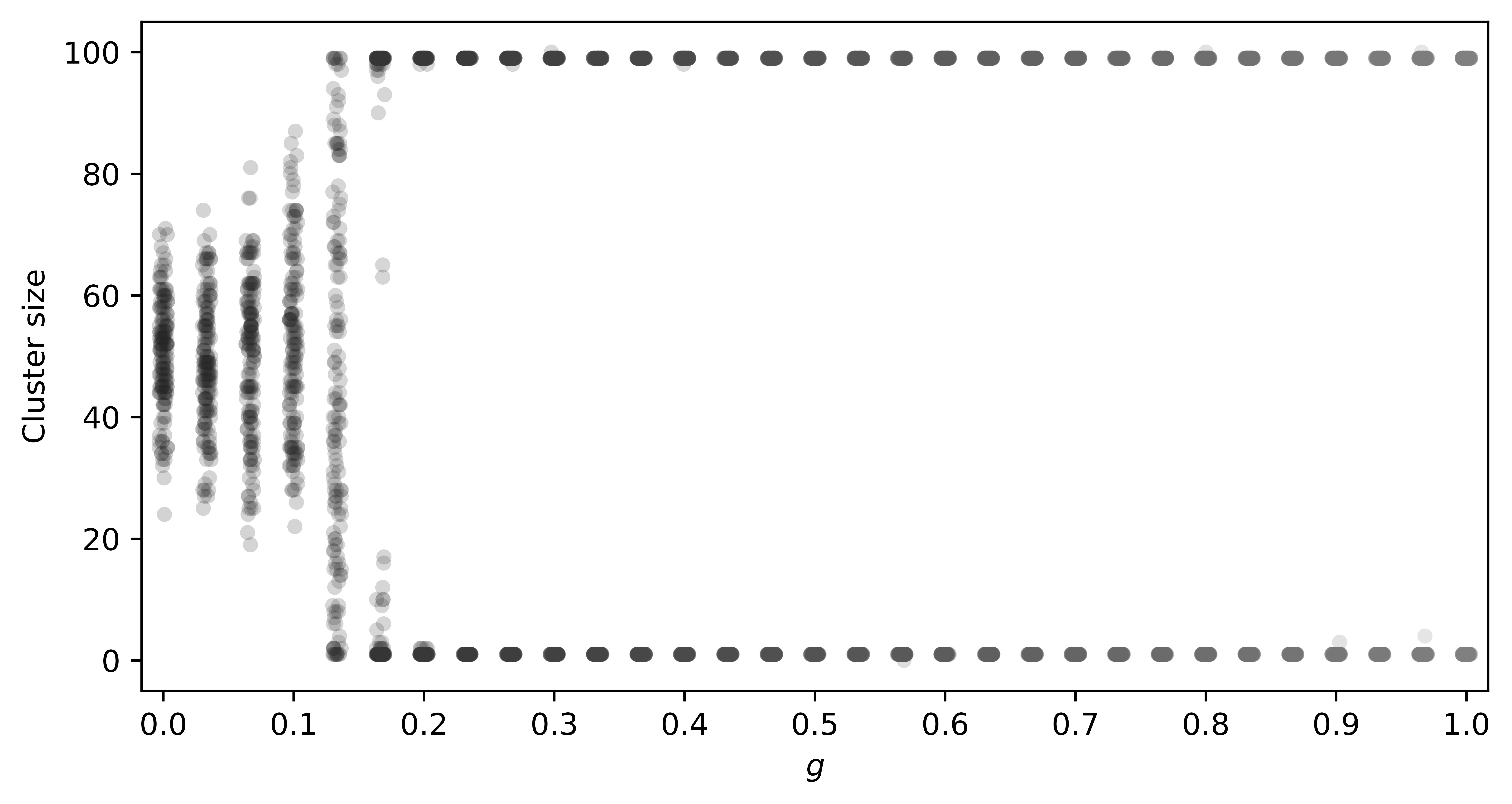}
		\caption{$d=150$}
		\label{fig:ScatterRandomClusterSizesD150}
	\end{subfigure}
	\caption{Scatterplot of cluster size resulting from WCI clustering of standard Gaussian samples with different choices of $g$ and $d$. Values of $g$ between 0 and 0.5 produce the widest range of cluster sizes.}
	\label{fig:WishbonePlots}
\end{figure}

The key feature of the WCI is that it does not favor balanced clusters as strongly as the CI.
We now show how this works directly by clustering simulated Gaussian data in several choices of dimension.
For each of $d \in \s{1, 4, 32, 100}$, and each $g$ in a grid on $[0,1]$, we simulate 150 iid samples of size 100 from $N_d(\bzero, \Ib_d)$, cluster them with WCI clustering, and plot the size of one of the two clusters chosen at random.
Figure \ref{fig:WishbonePlots} shows a plot for each choice of dimension $d$.
Each scatterplot has a ``wishbone'' shape: small values of $g$ produce relatively balanced clusters; as $g$ increases, a wider range of cluster sizes is produced; and eventually only \textit{unbalanced} clusters are produced.
The wishbone contracts horizontally toward the left margin as the dimension $d$ increases.
Note that a wide distribution of cluster sizes is produced just before the fork in the wishbone.
Depending on the dimension, this occurs somewhere between $g=0$ and 0.6.
The locations of the wishbones in Figure \ref{fig:WishbonePlots} depend both on the dimension and on the eigenvalues of Gaussian covariance.
However, for fixed $d$, equal eigenvalues produce the right-most wishbone. Therefore examples with unequal eigenvalues do not give additional insight.

In conclusion, the WCI is a clustering criterion that is like the CI but allows smaller clusters to play a greater role.
In particular, when the data are Gaussian, the WCI with appropriate choice of $g$ is more or less impartial to how balanced the cluster sizes are.

\subsection{Weighted SigClust}
\label{sec:sigclust:wcisigclust}
SigClust has low power when facing unbalanced clusters because its test statistic, the CI, is not sensitive to small clusters.
Therefore, to increase SigClust's power in this setting, we propose \textit{Weighted SigClust}, which replaces the test statistic with the WCI (\ref{eq:wci}), a criterion that \textit{is} sensitive to small clusters.
Any value of the tuning parameter $g$ in the WCI may be used; we recommend trying 0, 0.25, and 0.5 and using the choice with the strongest z-score.
In particular, Weighted SigClust compares the WCI of the clustering of interest to a null distribution formed by minimizing the WCI on simulated datasets drawn under $H_0$, as discussed in Section \ref{sec:sigclust:sigclustreview:sigclust}.
As in conventional SigClust, we reject $H_0$ if the sample WCI is small enough with respect to this null distribution.

\begin{figure}[h!]
	\centering
	\begin{subfigure}{.31\textwidth}
		\centering
		\includegraphics[width=\textwidth]{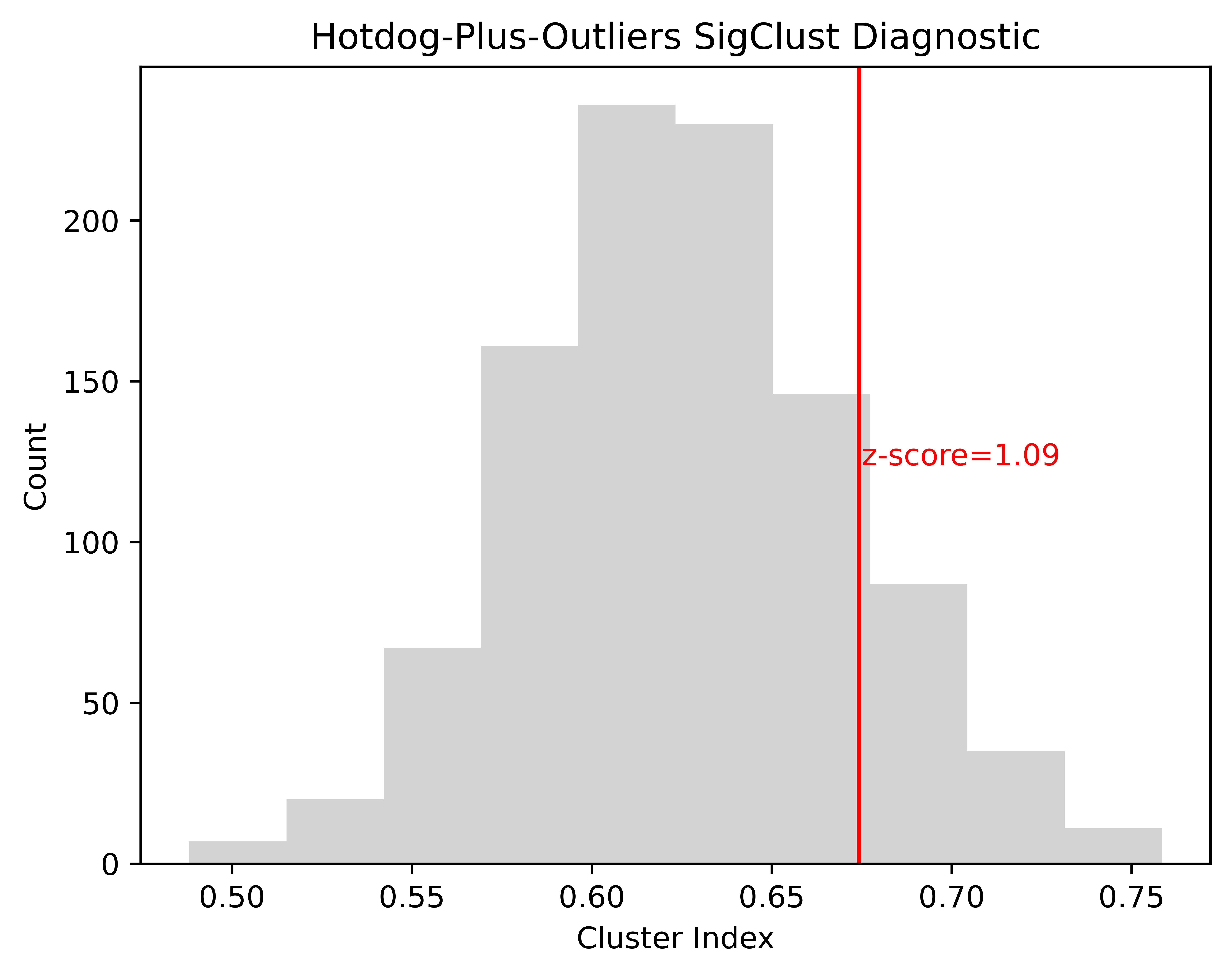}
		\caption{\textsmaller{Conventional SigClust}}
	\end{subfigure}
	\hfill
	\begin{subfigure}{.31\textwidth}
		\centering
		\includegraphics[width=\textwidth]{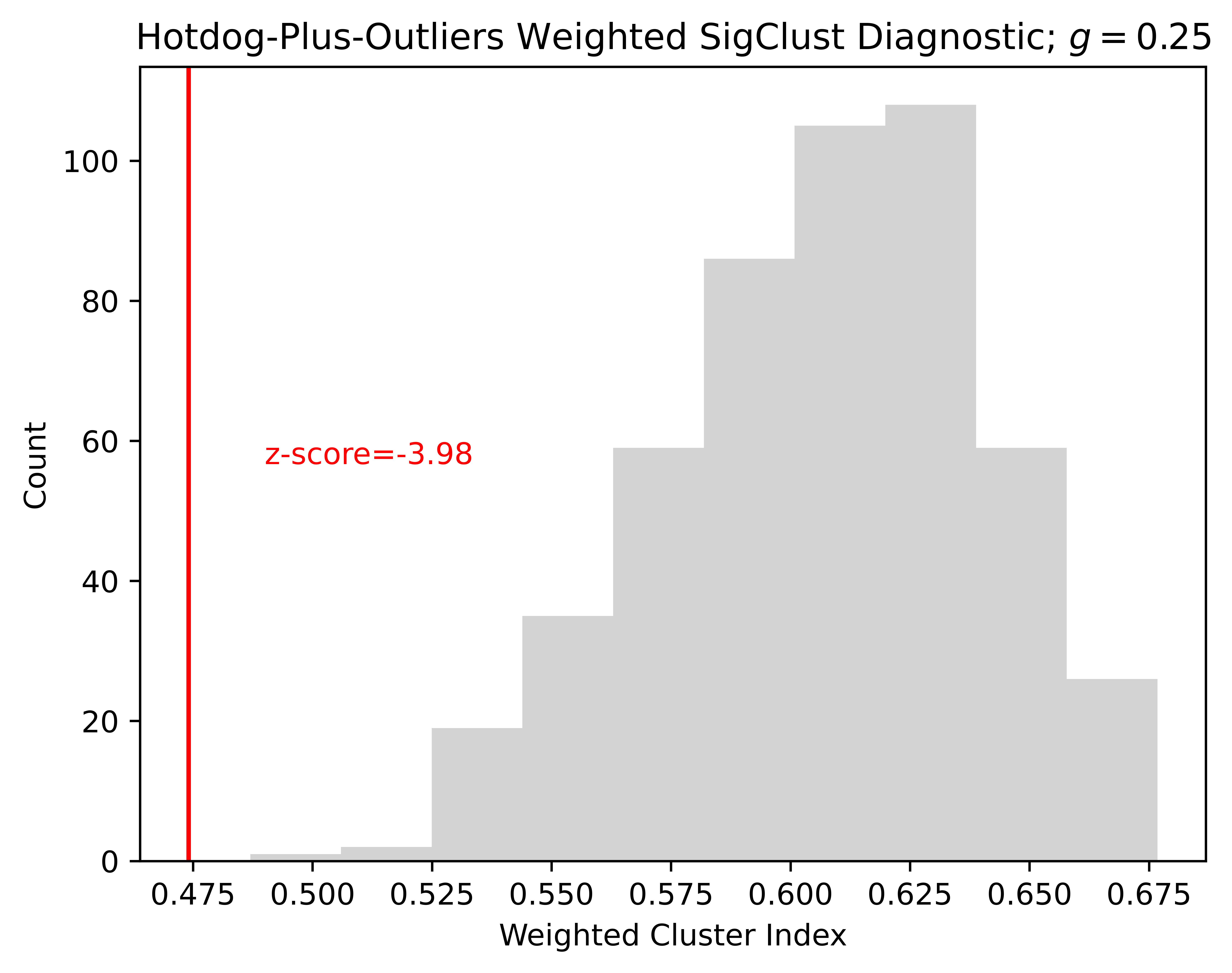}
		\caption{\textsmaller{Weighted SigClust, $g=0.25$}}
	\end{subfigure}
	\hfill
	\begin{subfigure}{.31\textwidth}
		\centering
		\includegraphics[width=\textwidth]{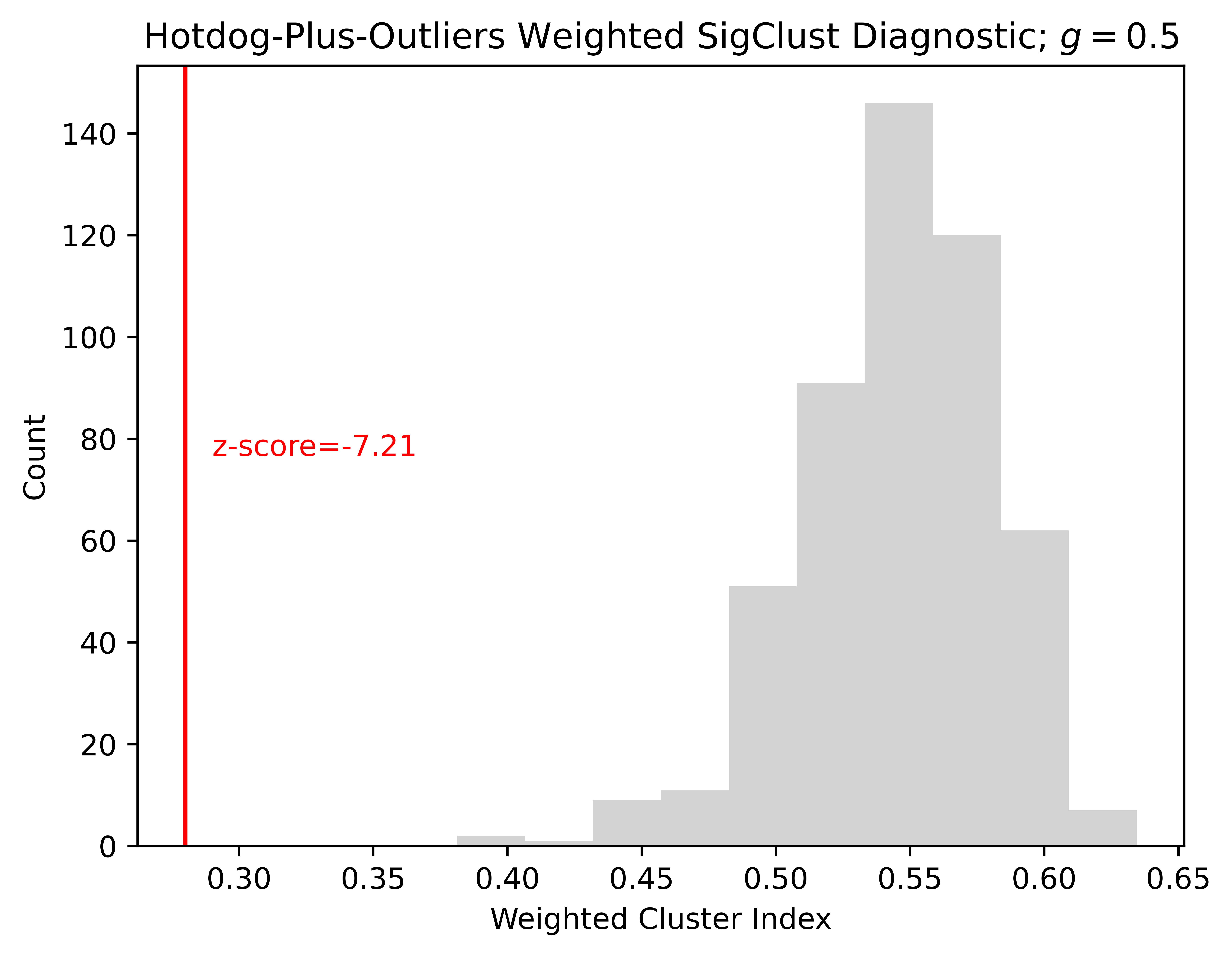}
		\caption{\textsmaller{Weighted SigClust, $g=0.5$}}
	\end{subfigure}
	\caption{Comparison of conventional and Weighted SigClust for Hotdog-Plus-Outliers data. Conventional SigClust cannot reject $H_0$, while the Weighted SigClust does, with a very strong z-score, for both $g=0.25$ and 0.5.}
	\label{fig:hotdog_plus_outliers_sigclusts}
\end{figure}

We now apply the proposed Weighted SigClust to the Hotdog-Plus-Outliers data introduced in Section \ref{sec:sigclust:wci}, and compare the result to the conventional SigClust.
The diagnostic plots are in Figure \ref{fig:hotdog_plus_outliers_sigclusts}.
We see that the conventional SigClust is not able to validate this clustering: the diagnostic shows that the sample CI is in the range of the typical values under the null hypothesis. This is quantified by the z-score of 1.09.
Weighted SigClust, on the other hand, finds decisive evidence of clustering with both $g=0.25$ and 0.5. The sample WCIs are smaller than any of the simulated null statistics; and the associated z-scores are very strong: -3.98 and -7.21 respectively.
In the Appendix we extend this example, providing Sigclust diagnostics for more values of $g$ along with plots of the distributions of the cluster sizes in the simulation as in the ``wishbone plots'' in Figure \ref{fig:WishbonePlots}.
We conjecture that power is maximized when these distributions are closest to uniform.

\begin{figure}[htbp]
	\centering
	\begin{subfigure}{.45\textwidth}
		\centering
		\includegraphics[width=\textwidth]{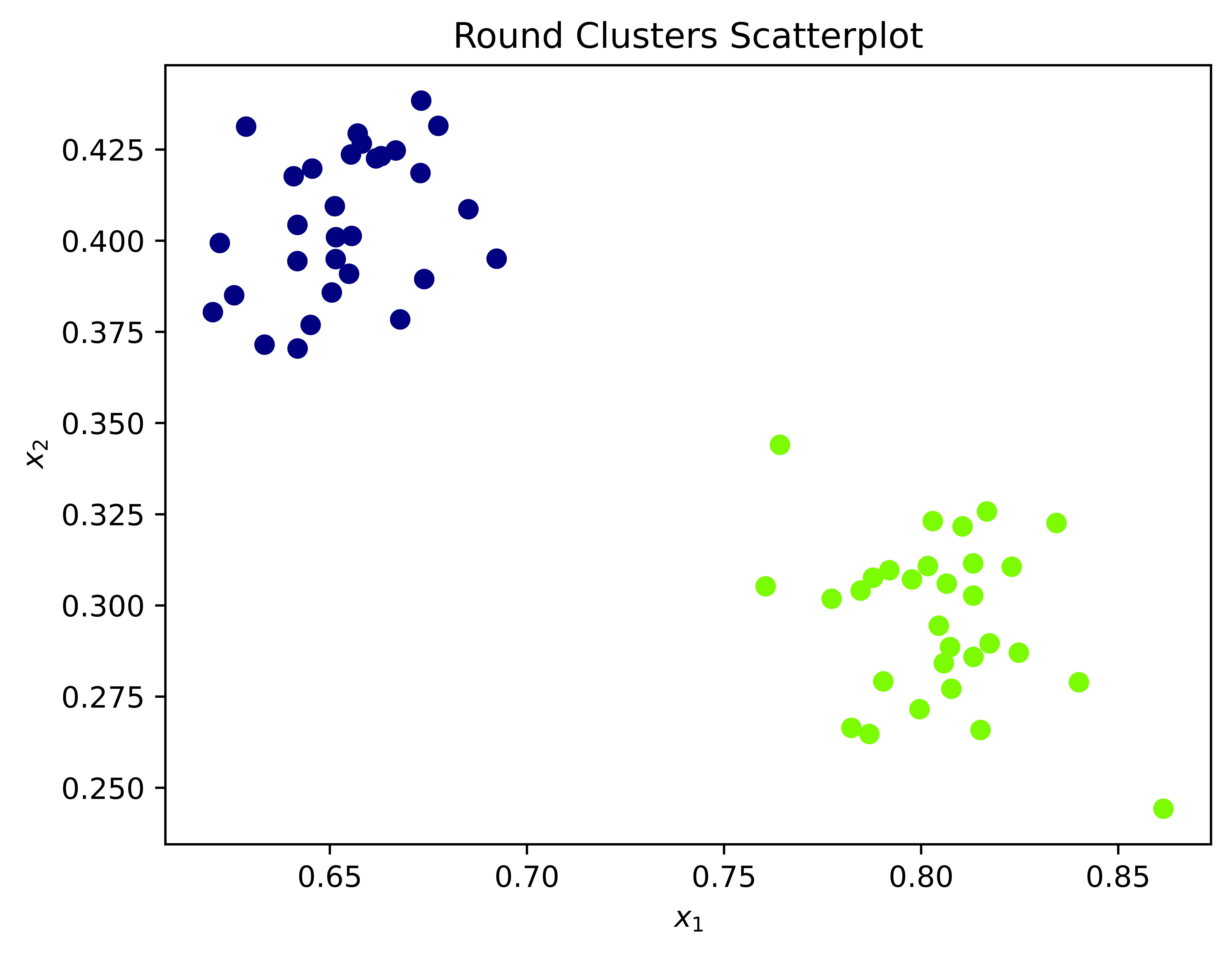}
		\caption{Round Clusters scatterplot.}
		\label{fig:round_clusters_scatter}
	\end{subfigure}
	\hfill
	\begin{subfigure}{.45\textwidth}
		\centering
		\includegraphics[width=\textwidth]{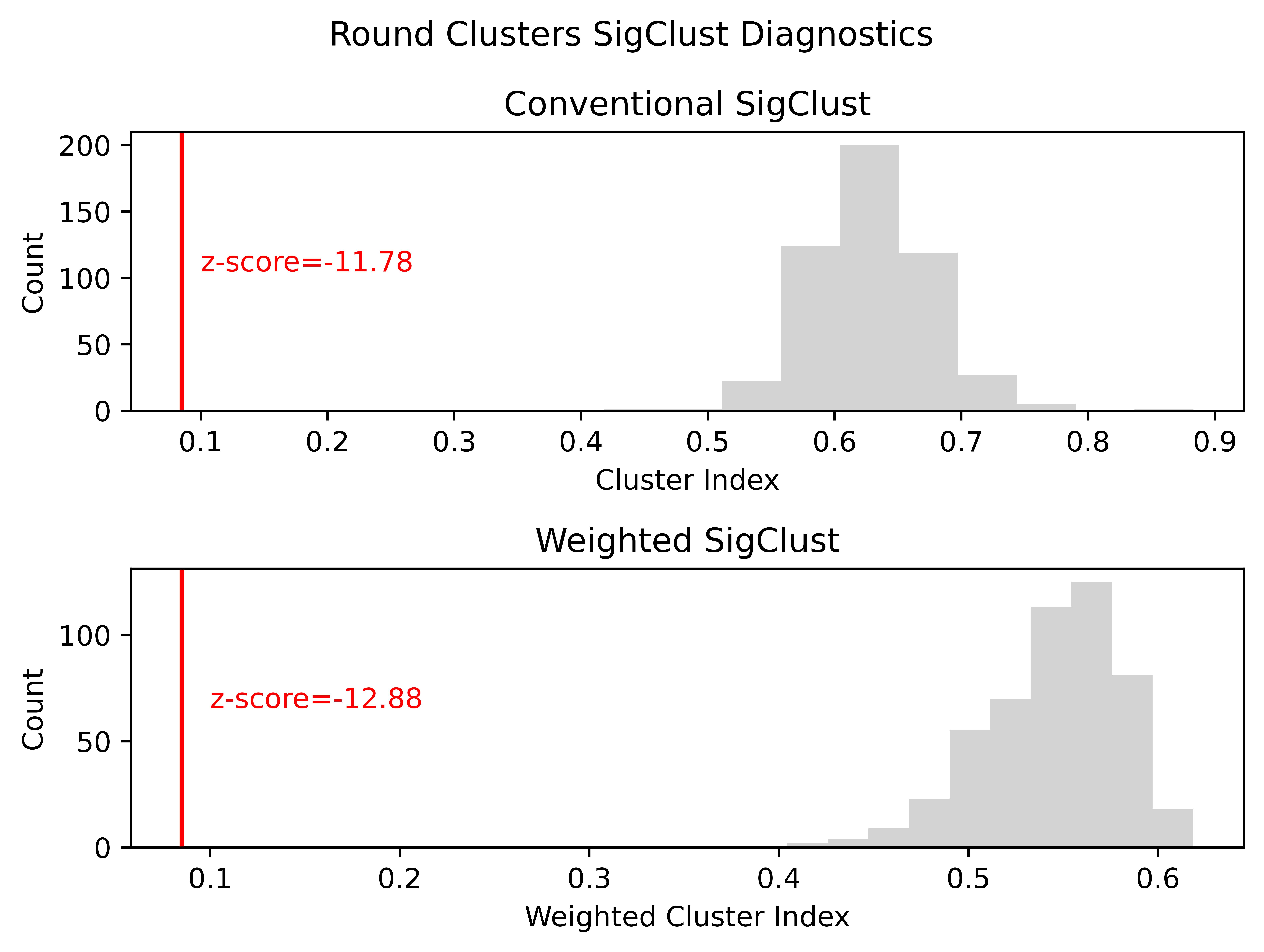}
		\caption{Conventional and Weighted SigClust diagnostics.}
		\label{fig:round_clusters_sigclusts}
	\end{subfigure}
	\caption{An example of balanced clusters. Both conventional and Weighted SigClust with $g=0.5$ recognize this clustering as strongly significant.}
	\label{fig:round_clusters}
\end{figure}

While Weighted SigClust is much more powerful than the conventional SigClust in the previous example, we also provide an example showing that our method does not \textit{lose} power in the balanced cluster setting.
In Figure \ref{fig:round_clusters_scatter} we show a scatterplot of two simulated round clusters, each of thirty points in $\R^2$.
Like Hotdog-Plus-Outliers, these data are a subset of the \textit{Four Clusters} dataset in \cite{oodabook}.
In Figure \ref{fig:round_clusters_sigclusts} we compare the diagnostic plots of Weighted SigClust with $g=0.5$ and conventional SigClust on this dataset.
Both test statistics are far below any of the simulated null statistics; indicating that both methods are very powerful on this balanced example. To save space, we do not show the plot for $g=0.25$; its visual impression is in between the two shown.

\begin{figure}[htbp]
	\centering
	\begin{subfigure}{.45\textwidth}
		\centering
		\includegraphics[width=\textwidth]{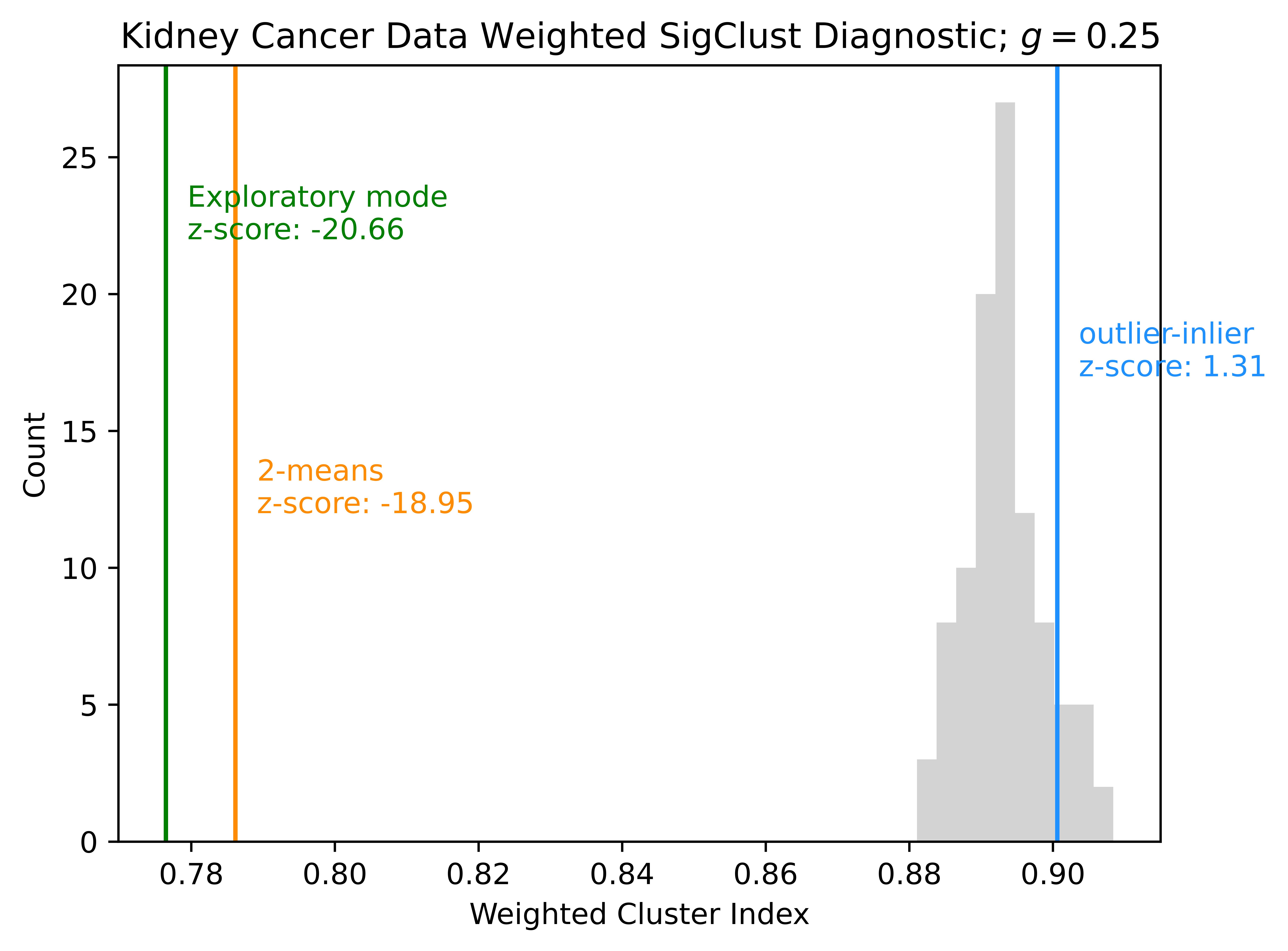}
		\caption{$g=0.25$}
	\end{subfigure}
	\hfill
	\begin{subfigure}{.45\textwidth}
		\centering
		\includegraphics[width=\textwidth]{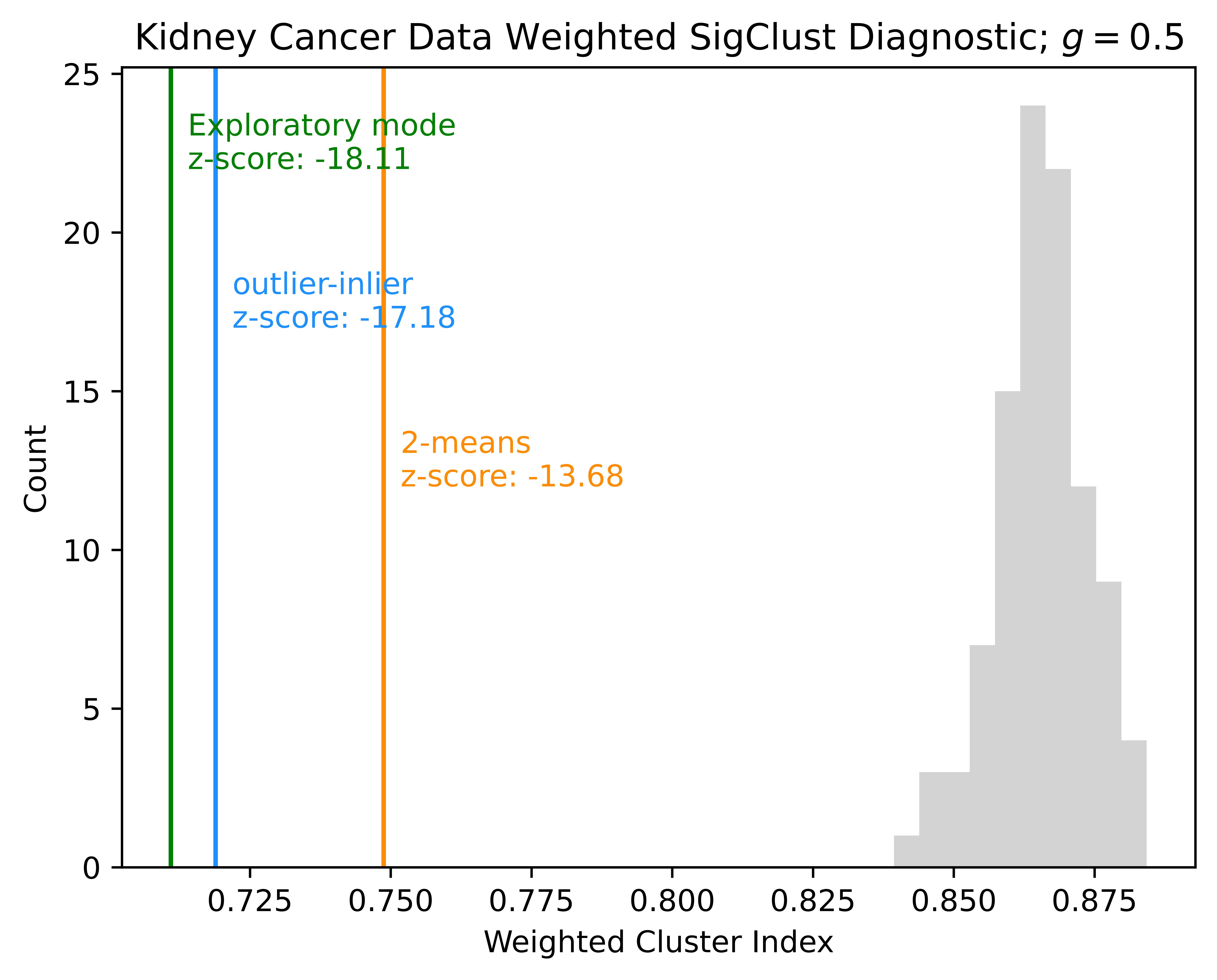}
		\caption{$g=0.5$}
	\end{subfigure}
	\caption{Weighted SigClust diagnostic plots for Kidney Cancer Data with $g=0.25$ and 0.5. With $g=0.5$, all three labelings provide strongly significant z-scores.}
	\label{fig:kidney_wci_sigclust}
\end{figure}

We now return to the Kidney Cancer Data from Section \ref{sec:sigclust:sigclustreview:sigclust}, and its two candidate labelings: \texttt{2-means} and \texttt{outlier-inlier}.
As was demonstrated in Figure \ref{fig:kidney_sigclust_conventional}, conventional SigClust finds strong support for \texttt{2-means} but not \texttt{outlier-inlier}.
However, using Weighted SigClust with $g=0.5$, we do find strong support for both labelings.
In Figure \ref{fig:kidney_wci_sigclust} we show the Weighted SigClust diagnostic plots for this dataset for $g=0.25$ and 0.5, with the WCIs and z-scores for both labelings.
For completeness, we also include the exploratory-mode WCI and z-score, associated with the labeling that minimizes WCI on the sample.
Not only has \texttt{outlier-inlier} become strongly significant using Weighted Sigclust with $g=0.25$, it is also more significant than \texttt{2-means}, and rivals the exploratory-mode WCI.

\section{Implementation details} \label{sec:sigclust:optimization}
In this section we discuss the optimization task of finding a clustering that minimizes the $\wci$.
The key difficulty is the combinatorial search space: there are $2^{n-1} -1$ ways to assign $n$ points to two clusters.
The $k$-means procedure is typically minimized using the iterative
algorithms of \cite{lloyd1957} or \cite{macqueen1967}.
These algorithms monotonically improve the objective with each iteration,
so they always find at least a local minimum of the CI.
However, this does not apply to the WCI, so a different approach is needed.

Our approach is to limit the $\bigO(2^n)$ search space to a $\bigO(n)$ space of partitions, where each partition is induced by a hyperplane normal to one of the top principal components of the data.
The idea is to slide a hyperplane along each of the top $P$ PCs and record the WCI for each of the resulting $P(n-1)$ partitions.
Figure \ref{fig:compare_cis_on_hotdog} provides a visual example of this scheme using the first principal component of the Hotdog-Plus-Outliers data.
Since a (near)-optimal partition might not be determined by PC1, the process may be repeated with PC2, 3, etc., taking the lowest overall WCI as the output.
The number $P$ of PCs with which to repeat this process is up to the user.
There is certainly no guarantee that the optimal clustering will be defined by a hyperplane normal to a top PC;
we therefore make two notes in defense of this approach.
First, in classical 2-means clustering, the optimal clusters can always be separated by a hyperplane.
We hypothesize that this is true for WCI clustering as well.
Second, among approaches using hyperplanes, defining the hyperplanes using the PC directions
can benefit from the variance structure of the data; this idea was also considered by
\cite{gapstat} in their development of the gap statistic.
The routine described above is straightforward to implement; pseudocode is found in Algorithm \ref{alg:slowpca}.
\begin{algorithm}[htbp]
	\caption{Algorithm to search PC $p$ for minimal $\wci$. \label{alg:slowpca}}
	\KwData{$\Xb \in \R^{d\times n}$ data to be clustered, $P$ number of principal components to search, $\Pb \in \R^{P \times n}$ scores matrix for top $P$ PCs, $g$ exponent to use for $\wci$.}
	$\mathbf{WCI} \in \R^{P \times (n-1)} \gets \bzero$, a matrix to hold candidate WCIs\;
	\For{$p = 1, \cdots, P$}{
		$\tilde{\Xb} \gets \Xb$ with observations sorted by PC $p$ score\;
		\For{$i=1,\cdots,n-1$}{
			$\mathrm{WCI}_{p, i} \gets \wci_g(\tilde{\Xb}_{:, 1:i}, \tilde{\Xb}_{:, (i+1):n})$\;
		}
	}
	\KwRet{$\min_{p, i} \mathrm{WCI}_{p, i}$}
\end{algorithm}
However, in practice we recommend using an accelerated version of this routine, which we describe next.

Accelerating Algorithm \ref{alg:slowpca} is accomplished by rewriting the within-cluster sum-of-squares terms as functions of the pairwise squared distances $d_{ij} = ||\bx_i-\bx_j||^2$, using the following well-known identity:

\begin{lemma}
	\label{pairwiselemma}
	Let $\bx_1, \cdots, \bx_n \in \R^d$, let
	$\bxbar = \frac{1}{n} \sum_{i=1}^n \bx_i$,
	and let $d_{ij} = ||\bx_i - \bx_j||^2$. Then
	\begin{equation}
	\sum_{i=1}^n ||\bx_i - \bxbar ||^2
	= \frac{1}{2n} \sum_{i, j = 1}^n d_{ij}.
	\end{equation}
\end{lemma}
This lemma will allow us to organize the calculations such that checking the next partition can be performed via an update of previous calculations.
This update step is what provides the acceleration.

To search along the $p$\textsuperscript{th} PC, we first define the following.
Let $d_{ij}$ and $\bxbar$ be defined as in Lemma
\ref{pairwiselemma} for the observed data $\s{\bx_1, \cdots, \bx_n}$,
and let $r_i = ||\bx_i - \bxbar||^2$ be the squared distances to the overall mean.
Sort the data indices by PC $p$ score, so that
$\pi(\cdot)$ represents the permutation of
$\s{1, \cdots, n}$ that sorts $\s{\bx_1, \cdots, \bx_n}$ by PC $p$ score.
Then, any two clusters formed by splitting along this PC will have sizes $k$ and $n-k$ for some $1 \leq k < n$, and can be written as
$C_1 = \s{\pi(1), \cdots, \pi(k)}$ and
$C_2 = \s{\pi(k+1), \cdots, \pi(n)}$.

The the $\wci$ for these clusters would be written
\begin{equation} \label{eq:cig_permutation_nolemma}
	\frac{
		k^{-0.5} \sum_{i=1}^k ||\bx_{\pi(i)} - \bxbar^\p{1}||^2
		+
		(n-k)^{-0.5} \sum_{i=k+1}^n ||\bx_{\pi(i)} - \bxbar^\p{2}||^2
	}
	{
		k^{-0.5} \sum_{i=1}^k r_{\pi(i)}
		+
		(n-k)^{-0.5} \sum_{i=k+1}^n r_{\pi(i)}
	}.
\end{equation}
Applying Lemma \ref{pairwiselemma} to the sums in the numerator, (\ref{eq:cig_permutation_nolemma}) becomes
\begin{equation} \label{eq:cig_permutation}
	\dfrac{
		2k^{-1.5} \sum_{i, j = 1}^{k} d_{\pi(i), \pi(j)}  +
		2(n-k)^{-1.5} \sum_{i, j = k+1}^{n} d_{\pi(i), \pi(j)}
	}
	{
		k^{-0.5} \sum_{i=1}^k r_{\pi(i)}
		+
		(n-k)^{-0.5} \sum_{i=k+1}^n r_{\pi(i)}
	}.
\end{equation}
For our updating algorithm, we will label the four summations in (\ref{eq:cig_permutation}):
\begin{align*}
	\alpha_k &= \sum_{i, j = 1}^{k} d_{\pi(i), \pi(j)} \\
	\beta_k &= \sum_{i, j = k+1}^{n} d_{\pi(i), \pi(j)} \\
	\gamma_k &= \sum_{i=1}^k r_{\pi(i)} \\
	\delta_k &= \sum_{i=k+1}^n r_{\pi(i)}.
\end{align*}
Then, (\ref{eq:cig_permutation}) is succinctly written
\begin{equation} \label{eq:cig_permutation_short}
	\dfrac{
		2k^{-1.5} \alpha_k +
		2(n-k)^{-1.5} \beta_k
	}
	{
		k^{-0.5} \gamma_k +
		(n-k)^{-0.5} \delta_k
	}.
\end{equation}
To check the next partition on this PC, we must find the $(k+1)$\textsuperscript{th}
values of $\alpha$, $\beta$, $\gamma$, $\delta$.
Next we will show that these values can be specified in terms of an efficient update of their $k$\textsuperscript{th} values.

We explain the update step visually using Figure \ref{fig:Dmatrix}, which
shows the matrix $\s{d_{ij}}$ with rows and columns ordered by $\pi(\cdot)$.
\begin{figure}[htbp]
	\centering
	\[
	\begin{bNiceMatrix}[margin,first-row, first-col]
		& 1 & & & & k & & & n\\
		1 & \Block[draw=orange,rounded-corners, line-width=2pt]{5-5}{} 0 & \cdot & \cdot & \cdot & \cdot & \Block[fill=orange!15, rounded-corners]{5-1}{} \cdot & \cdot & \cdot \\
		& \cdot & 0     & \cdot & \cdot & \cdot & \cdot & \cdot & \cdot \\
		& \cdot & \cdot & 0     & \cdot & \cdot & \cdot & \cdot & \cdot \\
		& \cdot & \cdot & \cdot & 0     & \cdot & \cdot & \cdot & \cdot \\
		k & \cdot & \cdot & \cdot & \cdot & 0     & \cdot & \cdot & \cdot \\
		& \Block[fill=orange!15, rounded-corners]{1-5}{} \cdot & \cdot & \cdot & \cdot & \cdot & \Block[draw=blue, rounded-corners, line-width=2pt]{3-3}{} 0   & \Block[fill=blue!15,rounded-corners]{1-2}{} \cdot & \cdot \\
		& \cdot & \cdot & \cdot & \cdot & \cdot & \Block[fill=blue!15,rounded-corners]{2-1}{} \cdot & 0   & \cdot \\
		n & \cdot & \cdot & \cdot & \cdot & \cdot & \cdot & \cdot & 0
	\end{bNiceMatrix}
	\]
	\caption{Distance matrix with rows and columns ordered by PC score, illustrating the updating step of the algorithm.
	}
	\label{fig:Dmatrix}
\end{figure}
The values of $\alpha_k$ and $\beta_k$ are
given by the sums of the values in the orange and blue-bordered boxes respectively.
We can see that $\alpha_{k+1}$ and $\beta_{k+1}$ would be found by expanding the orange-bordered box to include the orange-shaded values while shrinking the blue-bordered box to exclude the blue-shaded values.
Therefore, to perform the update,
we need only add the terms shaded in orange to $\alpha_k$,
and subtract the terms shaded in blue from $\beta_k$.
The values of $\gamma_k$ and $\delta_k$ are updated by adding $r_{\pi(k+1)}$ to the former and subtracting it from the latter.
The pseudocode to search along PC $p$ is given in Algorithm \ref{alg:fastpca}.

\begin{algorithm}[htbp]
	\caption{Updating algorithm to search PC $p$ for minimal $\wci$. \label{alg:fastpca}}
	\newcommand{\pluseq}{\mathrel{+}\mathrel{\mkern-1mu}=}
	\newcommand{\minuseq}{\mathrel{-}\mathrel{\mkern-1mu}=}
	\KwData{$\Db = \s{d_{ij}}$, $\br = \s{r_i}$, $\pi(\cdot)$.}
	$\mathbf{WCI} \in \R^{n-1} \gets \bzero$, a vector to hold candidate CIs\;
	$\Db^\pi \gets \Db$ with rows and columns ordered by $\pi(1), \cdots, \pi(n)$\;
	$\alpha_0 \gets 0$\;
	$\beta_0 \gets \sum_{i, j} d_{ij}$\;
	$\gamma_0 \gets 0$\;
	$\delta_0 \gets \sum_{i=1}^n r_i$\;
	\For{$k = 0, \cdots, n-2$}{
		$\alpha_{k+1} \gets \alpha_k + 2 \sum_{j=1}^{k} D^\pi_{k+1, j}$ i.e., add the orange shaded values in Figure \ref{fig:Dmatrix}\;
		$\beta_{k+1} \gets \beta_k - 2 \sum_{j=k+2}^n D^\pi_{k+1, j}$, i.e., subtract the blue shaded values in Figure \ref{fig:Dmatrix}\;
		$\gamma_{k+1} \gets \gamma_k + r_{\pi(k)}$\;
		$\delta_{k+1} \gets \delta_k - r_{\pi(k)}$\;
		$\wci_{k+1} \gets \dfrac{
			2(k+1)^{-1.5} \alpha_{k+1} +
			2(n-k-1)^{-1.5} \beta_{k+1}
		}
		{
			(k+1)^{-0.5} \gamma_{k+1}+
			(n-k-1)^{-0.5} \delta_{k+1}
		}$\;
	}
	\KwRet{$\min_{1 \leq k \leq n-1 } \wci_{k}$}
\end{algorithm}

\section{Discussion} \label{sec:sigclust:discussion}
The discussion section collects several elements.
Section \ref{sec:rigorissues}
points out and resolves a potential lack of statistical rigor concerning confirmatory-mode SigClust.
Section \ref{sec:additionaldisco} remarks on a recent asymptotic study of SigClust and concludes the paper.

\subsection{Rigorous statistical inference issues in SigClust}
\label{sec:rigorissues}
We raise a fine statistical point about the soundness of SigClust that has not yet been addressed in the literature: SigClust is only a valid hypothesis test when both the sample and the synthetic datasets are clustered using the same procedure, that is, when SigClust is used in exploratory mode.
Here we discuss why this is the case and offer a resolution.

In exploratory mode (recall Definition \ref{def:sigclustmodes}),
the same clustering procedure is used to cluster both the sample \textit{and} the synthetic datasets used in the null distribution.
In conventional SigClust this is 2-means clustering, and in Weighted SigClust it is WCI clustering.
In confirmatory mode, by contrast, the procedure used to generate the candidate labeling may not be the same as the one used for generating the null CIs.
Hence, the labeling may not provide the minimal achievable CI on the sample.
In fact, the candidate labeling is often produced by visual inspection, as with the \texttt{outlier-inlier} labels, or another ad hoc procedure that cannot be specified as a pure function of the data.
This issue can be better examined by rigorously defining SigClust's test statistic.
Without loss of generality, we discuss conventional SigClust; the same discussion applies to Weighted SigClust, substituting the WCI for the CI.
Recall that $\mathcal{P}_n$ denotes the binary partitions of $\s{1, \cdots, n}$.
In exploratory mode, we use the sample $\Xb$ to produce the test statistic
$T(\Xb) := \min_{(C_1, C_2) \in \mathcal{P}_n} \ci(C_1, C_2)$, and the distribution of $T$ under $H_0$ can be estimated using the simulation procedure described in Section \ref{sec:sigclust:sigclustreview:sigclust}.
However, in confirmatory mode, the sample uses
$\tilde{T}(\Xb, \bm{f}) := \ci( \bm{f(\mathrm{X})} )$,
where
$\bm{f}: \R^{n \times d} \to \mathcal{P}_n$ is the clustering procedure used to generate the candidate labels, but the CIs in the null distribution
are still produced by $T$, not $\tilde{T}$.
This means that, in the sense of strict mathematical statistics, SigClust is only a valid hypothesis test when used in exploratory mode, because the sample and the null use the same test statistic.
A rigorous recipe for SigClust might be to specify the clustering routine $\bm{f}$ in advance,
and use that same routine for both the sample and the null distribution.

However, this discussion does not invalidate the use of SigClust in confirmatory mode, as we will show using the kidney cancer example.
The diagnostic plot in Figure \ref{fig:kidney_wci_sigclust},
shows WCIs for three labelings: the exploratory mode labeling, i.e., the labeling that minimizes the WCI statistic, the \texttt{outlier-inlier} labeling, and the \texttt{2-means} labeling.
The latter two are confirmatory-mode uses of SigClust.
The diagnostic plot shows that
all three WCIs are far smaller than the WCIs simulated under $H_0$.
Rejecting $H_0$ for the exploratory-mode CI is absolutely justified, because the same clustering procedure was used for both the sample and the synthetic datasets.
On the other hand, the \texttt{outlier-inlier} labeling came from visual inspection of the PCA scatterplot in Figure \ref{fig:kidney_pca}, and the \texttt{2-means} labels came from minimizing the CI, not the WCI.
The discussion above would suggest that rejecting $H_0$ for these labelings would not be statistically valid because the same function was not used for both the sample and the synthetic datasets.
We offer the following resolution.
The WCIs simulated under the null are explicitly trying to minimize the criterion.
The WCIs of our candidate labelings are smaller than all of these simulated ones, without even trying to be minimal.
\textit{They are both therefore much stronger clusterings
with respect to WCI than the strongest that could be found under the null.}
This is exactly the sort of conclusion that SigClust is intended to provide.
We are therefore justified in using SigClust in confirmatory mode, in spite of the subtle issue discussed above.

\subsection{Additional issues}
\label{sec:additionaldisco}
The original SigClust paper
\citep{sigclust1} described a parametric bootstrap method to estimate the null distribution of the CI,
but recently, \cite{rift} proved an $n\to\infty$ central limit theorem for this distribution.
This raises the possibility of using asymptotic arguments in SigClust instead of simulation, which could provide a big improvement in speed, especially in high-dimensional settings.
We compared their CLT to results from simulation, and found that the sample size needs to be in the tens of thousands before the asymptotic approximation begins to match the simulated distribution.
Details will appear in the forthcoming dissertation \citep{KeefeDiss}.

An important question, then, is whether the rate of convergence in this CLT can be improved, because SigClust is often used with much smaller samples.
Furthermore, we conjecture that a similar CLT would apply to WCI as well.
If so, it would be helpful to understand its asymptotics not only as $n$ grows, but also as the clusters become more imbalanced, i.e., $|C_1| / |C_2| \to 0$.
Such a CLT would be useful for large sample sizes with highly imbalanced clusters:
the large $n$ would make simulation slow, and the imbalanced clusters would mean that subsampling to a more manageable $n$ would lose too much signal from the minority cluster.

We believe there are much better ways to minimize the WCI criterion.
Since it is hard to minimize the WCI globally, i.e., over all binary partitions,
we have chosen instead to minimize it over a particular subset of partitions using the ``sliding hyperplane'' scheme in Section \ref{sec:sigclust:optimization}.
Minimizing over this subset already provides a dramatic increase in SigClust's power in unbalanced examples,
but we would prefer a more principled approach, as well as one that could readily extend to $k > 2$.
The forthcoming dissertation \citep{KeefeDiss} will examine an approach based on semidefinite programming.

An open theoretical question involves the relationship between the power of the test and the sizes of the clusters that SigClust produces when simulating the null distribution of WCI.
This is further discussed in the Appendix.
In our examples, it seems that power is maximized when the distribution of cluster sizes is uniform; see, e.g., Figure \ref{fig:power_analysis_hotdog}.
This supports the idea that being more impartial to cluster size is what gives our method power in unbalanced settings.

\appendix
\appendixpage

\section{Extended Hotdog-Plus-Outliers SigClust example}
\begin{figure}[htb]
	\centering
	\includegraphics[width=8cm]{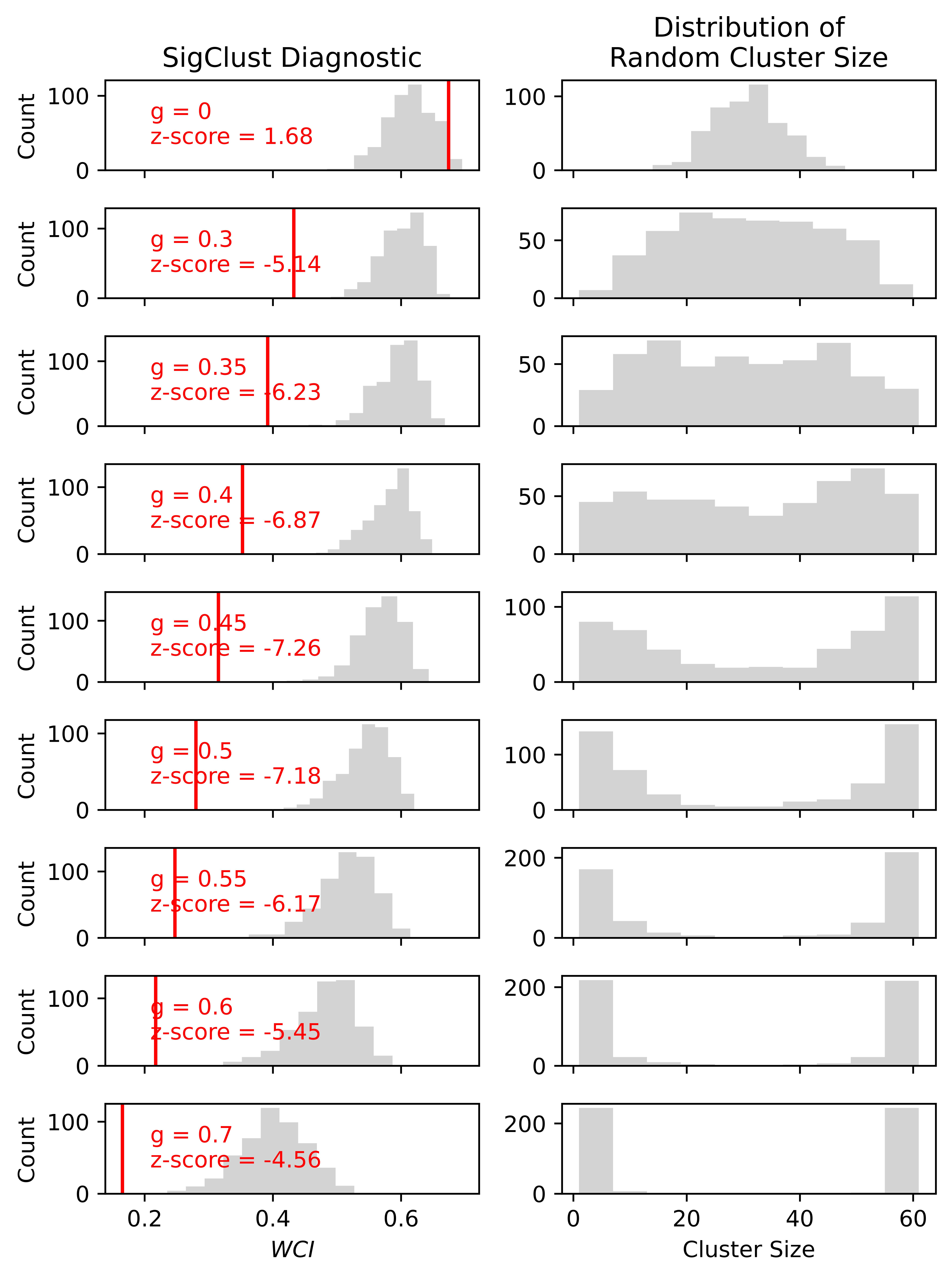}
	\caption{Weighted SigClust results for Hotdog-Plus-Outliers dataset using different exponents $g$. The column of SigClust diagnostic plots on the left shows that $g\in [0.3, 0.7]$ produce high power against $H_0$ for this example. The histograms on the right show that the clusters in the SigClust null simulation tend to be balanced for $g=0$, very unbalanced for $g \geq 0.6$, and exhibit a range of balances when $g \in [0.3,0.5]$.}
	\label{fig:power_analysis_hotdog}
\end{figure}

Here we devote more discussion to the Hotdog-Plus-Outliers example, whose SigClust diagnostic plots are in Figure \ref{fig:hotdog_plus_outliers_sigclusts}.
In Figure \ref{fig:power_analysis_hotdog} we collect Sigclust diagnostic plots for more values of $g \in [0, 0.7]$ (left), and histograms of the cluster sizes produced in the null simulation (right).
Each row of the figure corresponds to a choice of $g$. The left column contains the SigClust diagnostic plot, comparing the $\wci_g$ of the sample (red line) to the null distribution of $\wci_g$s (gray histogram). As in the previous SigClust diagnostic plots, we annotate the plot with the z-score as well.
The right column of the figure shows the distribution of the size  of the simulated clusters produced within each of the parametric bootstrap samples during the SigClust simulation.
In particular, for each bootstrap sample, one of the two clusters is chosen at random and its size is recorded in the histogram.
This distribution is comparable to the distributions visualized in the ``wishbone plots'' in Figure \ref{fig:WishbonePlots}.
This tells us whether the simulated clusters in the SigClust simulation tended to be balanced or unbalanced.
For example, in the top-right plot, we see that when $g=0$, the clusters tended to be roughly balanced, roughly 1:2 to 1:1.
For $g=0.4$ on the other hand, a wide range of balances were seen. For $g\geq 0.55$ the clusters were always very unbalanced, producing bimodal histrograms.

We have two takeaways from this figure.
From the left column, we see that values of $g \in [0.3, 0.7]$ all produce very negative z-scores, indicating high power against $H_0$ for this example.
In fact, $g=0.5$ gives the highest power, although this is not the case for all datasets.
From the right column we see that for $g=0$ the clustering method is very partial to balanced clusters, and for $g \geq 0.6$ very partial to unbalanced clusters. We conjecture that high power tends to be produced when the cluster size distribution in the right-column histograms is close to uniform. In this example, which here occurs for $g$ around $0.4$.

Following our conjecture above,
we pose the open problem of finding a \textit{Gaussian-impartial} clustering method,
which we define as a clustering method that would produce a uniform distribution of cluster sizes given Gaussian data.
The following definition makes this precise.

\begin{definition}[Gaussian impartial]
	Let $T_n: (\R^{d\times n} \times \mathcal{P}_n) \to \R$ be a (binary) clustering criterion, recalling that $\mathcal{P}_n$ denotes the two-set partitions of $\s{1, \cdots, n}$.
	Let $f_n: \R^{d \times n} \to \mathcal{P}_n$ be the clustering method that seeks to minimize $T_n$, i.e.,
	\begin{equation}
		f_n(\Xb) = \argmin_{(C_1, C_2) \in \mathcal{P}_n} \; T_n(\Xb, (C_1, C_2)).
	\end{equation}

	Let $X_1, X_2, \cdots \in \R_d$ be drawn iid from $N(\bmu, \bm{\Sigma})$ for some mean $\mu$ and covariance $\bm{\Sigma}$, and let $\Xb_n = [X_1, \cdots, X_n]^T$.
	Let $C$ be a cluster chosen uniformly at random from the two in $f_n(\Xb_n)$, and independently of $\Xb_n$.
	Then we say $f_n$ is \textit{Gaussian impartial with respect to $d, n, \bmu, \bm{\Sigma}$} if the distribution of $|C|$ is the uniform distribution on $\s{1, \cdots, n-1}$.
\end{definition}

A Gaussian-impartial clustering method would be even more impartial about where to split the Gaussian measure than the criteria in this work, and therefore should exhibit \textit{no} bias toward balanced or unbalanced clusters.
Such a method could be even more powerful against alternatives of small clusters than the WCI method presented in this work.

\section*{Funding}
The authors gratefully acknowledge funding from NSF DMS-2113404 and NIAMS P30AR072580.

\section*{Disclosure Statement}
The authors report there are no competing interests to declare.

\bibliographystyle{abbrvnat}
\bibliography{references.bib}

\end{document}